\documentclass[review]{elsarticle}
\usepackage{mathtools}
\usepackage{comment}
\usepackage{float}
\usepackage{multirow}
\usepackage{amsthm}
\usepackage{graphicx}
\usepackage[export]{adjustbox}
\usepackage{changepage}
\usepackage{url}
\usepackage{amssymb}
\usepackage{ulem}
\usepackage{array}
\usepackage{booktabs}
\usepackage[table]{xcolor}
\usepackage{hyperref}
\usepackage{mdframed}
\usepackage{rotating}

\hypersetup
{
  colorlinks   = true, %Colours links instead of ugly boxes
  urlcolor     = red,  %Colour for external hyperlinks
  linkcolor    = blue, %Colour of internal links
  citecolor    = blue  %Colour of citations
}
\usepackage{wasysym}
\usepackage{listings}
\usepackage{pdflscape}
\usepackage{rotating}
\usepackage[utf8]{inputenc}
\usepackage[T1]{fontenc}
\usepackage{tikz,lipsum,lmodern}
\usepackage[most]{tcolorbox}
\usepackage{pgfplots}
\usepackage[toc,page]{appendix}
\usepackage{textcomp}
\usepackage{epstopdf}
\usepackage{csquotes}
\usepackage{dirtytalk}
\usepackage{parskip}
\usepackage{amssymb}
\usepackage{tikz}
\usepackage{comment}
\usepackage{indentfirst}% this helps to indent the first paragraph after the heading or chapter
\usepackage[colorinlistoftodos,prependcaption,textsize=tiny]{todonotes}
\usepackage[english]{babel}
\usepackage{geometry}
\pgfplotsset{compat=1.9}
\usepackage{lineno}
\usepackage{caption} 
\usepackage{adjustbox}
\usepackage{pdfpages}
 \usepackage{ragged2e} 
\usepackage{xurl}
\usepackage{longtable,tabu}
\usepackage{afterpage}
\usepackage{verbatim}
\usepackage{subfigure}
\usepackage{subfloat}
\usepackage{longtable}
\usepackage{ltxtable}
\usepackage{fancyvrb}
\usepackage{tabularx}
\usepackage{makecell}
\usepackage{supertabular}

\journal{Elsevier}

\begin{document}
%\begin{sloppypar}
\begin{frontmatter}

\title{Assessing UML Diagrams by \textcolor{black}{GPT}: Implications for Education}
    
%\author{Beian Wang}
%\ead{wba20021119@whu.edu.cn}
%\author{Chong Wang}
%\ead{cwang@whu.edu.cn}
%\author{Peng Liang}
%\ead{liangp@whu.edu.cn}
%\author{Jie Liang}
%\ead{liangjie@wsyu.edu.cn}
%\address{School of Computer Science, Wuhan University, 430072 Wuhan, China}
%\address{School of Information Science and Engineering, Wuchang Shouyi University, Wuhan 430062, China}

\author[mymainaddress]{Chong Wang\corref{equalcontribution}}
\cortext[equalcontribution]{The two authors have equal contribution.}
\ead{cwang@whu.edu.cn}
\address[mymainaddress]{School of Computer Science, Wuhan University, 430072 Wuhan, China}

\author[mymainaddress]{Beian Wang\corref{equalcontribution}}
\ead{wba20021119@whu.edu.cn}

\author[mymainaddress]{Peng Liang\corref{mycorrespondingauthor}}
\cortext[mycorrespondingauthor]{Corresponding author at: School of Computer Science, Wuhan University, China. Tel.: +86 27 68776137; fax: +86 27 68776027.}
\ead{liangp@whu.edu.cn}

\author[mysecondaryaddress]{Jie Liang}
\ead{liangjie@wsyu.edu.cn}
\address[mysecondaryaddress]{School of Information Science and Engineering, Wuchang Shouyi University, Wuhan 430062, China}

\begin{abstract}
In software engineering (SE) research and practice, UML is well known as an essential modeling methodology for requirements analysis and software modeling in both academia and industry. In particular, fundamental knowledge of UML modeling and practice in creating high-quality UML diagrams are included in SE-relevant courses in the undergraduate programs of many universities. This leads to a time-consuming and labor-intensive task for educators to review and grade a large number of UML diagrams created by the students. Recent advances in generative AI techniques, such as \textcolor{black}{GPT}, have paved new ways to automate many SE tasks. However, current research or tools seldom explore the capabilities of \textcolor{black}{GPT} in evaluating the quality of UML diagrams. This paper aims to investigate the feasibility and \textcolor{black}{performance of GPT} in assessing the quality of UML use case diagrams, class diagrams, and sequence diagrams. First, 11 evaluation criteria with grading details were proposed for these UML diagrams. Next, a series of experiments was designed and conducted on 40 students' UML modeling reports to explore the performance of GPT in evaluating and grading these UML diagrams. The research findings reveal that GPT can complete this assessment task, but it cannot replace human experts yet. Meanwhile, there are five evaluation discrepancies between GPT and human experts. These discrepancies vary in the use of different evaluation criteria in different types of UML diagrams, presenting GPT's strengths and weaknesses in this automatic evaluation task. 
\end{abstract}

\begin{keyword}
UML Diagram, \textcolor{black}{GPT}, Model Assessment, Software Modeling Education
\end{keyword}

\end{frontmatter}

\section{Introduction}\label{chap:intro}
In software engineering (SE) research and practice, the Unified Modeling Language (UML) has been widely used in requirements engineering, software design, and system modeling by providing a set of standard graphic modeling notations~\citep{pilone2005uml}. For software engineers and developers, well-designed and structured UML diagrams can effectively bridge the gap between business requirements and technical implementation in software analysis and design. To help our students develop their competencies for future SE-related jobs, essential knowledge of UML modeling and practice on how to create high-quality UML diagrams are included in the SE-relevant courses of the undergraduate programs of many universities, such as \textit{Software Engineering}, \textit{Software Requirements Modeling}, \textit{Software Architecture Design}, etc. When teaching UML modeling in these courses, teachers normally offer an introduction to the most useful and standard UML diagrams, supplemented with UML modeling tasks for specified case studies. Generally, the UML diagrams created by the students are manually reviewed and evaluated by the teachers or teaching assistants to examine whether the students can correctly understand and use UML. However, it is a time-consuming and labor-intensive task for educators. 

Recent advancements in artificial intelligence (AI) techniques, especially the development of generative AI techniques, have paved new ways to automate many tasks in the software development life cycle. 
As a typical and widely used type of generative AI model, \textcolor{black}{GPT} has demonstrated remarkable capabilities in requirements analysis and modeling~\citep{combemale2023large}~\citep{luitel2024improving}~\citep{Combemale2023ChatGPT} with UML~\citep{Wang2024how}, software architecting~\citep{ahmad2023towards}, code generation~\citep{jiang2024survey}, etc. Meanwhile, many open-source tools have been developed for modeling education, as reported in \citep{agner2017survey}. However, current research or tools seldom explore the capabilities of GPT, or other generative AI models, in assessing the quality of software artifacts, i.e., the UML diagrams created by the students in this paper, acting as teachers to grade students' assignments.   

Therefore, this paper intends to investigate the feasibility and \textcolor{black}{performance of GPT} in assessing the quality of UML diagrams in SE education. First, according to the selected case study and the corresponding suggested answer, 11 evaluation criteria with grading details were proposed for three types of UML diagrams, i.e., UML use case diagrams, class diagrams, and sequence diagrams. Next, a series of experiments was designed and conducted on 40 students' UML modeling reports to explore the performance of GPT in evaluating UML diagrams with the proposed evaluation criteria. The results show that GPT can complete this assessment task, but it cannot replace human experts yet. Meanwhile, there are five evaluation discrepancies between GPT and human experts. These discrepancies vary in the use of different evaluation criteria in different types of UML diagrams, presenting GPT's strengths and weaknesses in this automatic evaluation task. 

The main \textbf{contributions} of our work are as follows:
\begin{itemize}
    \item A set of evaluation criteria is proposed for GPT to quantitatively assess the quality of three types of UML diagrams, i.e., use case diagrams, class diagrams, and sequence diagrams.
    \item An empirical study on 40 students’ UML modeling reports was conducted to synthesize five types of evaluation discrepancies and elaborate on the difference in GPT's competency when grading the three types of UML diagrams. The replication package for this work is available at~\citep{replpack}.
\end{itemize}

The remainder of this paper is organized as follows. Section 2 provides a background on UML and the principles of software modeling. Section 3 outlines the methodology used in our study, including research data, evaluation criteria for UML diagrams, evaluating processes conducted by GPT and human experts, etc. Section 4 presents two research questions and the corresponding results. Section 5 discusses the research findings and implications. Section 6 addresses the limitations of this work, followed by the conclusions and future work in Section 7.

\section{Related Work}\label{RelatedWork}
\label{chap:relat}
The recent advances in Large Language Models (LLMs) have ushered in a significant transformation in how we tackle numerous computational tasks, extending their impact from natural language processing to diverse domains. In the domain of SE, LLMs are increasingly proving their potential, particularly in automating routine tasks and enhancing both the precision and speed of software development processes~\citep{hou2023large}. 

\textcolor{black}{As reported in \citep{gao2025current}, `Comprehensive evaluation of generated requirements and design' and `Inconsistency between software modeling and natural language descriptions' are reported as two main challenges in Software Requirements and Design.} Especially in the context of Requirements Engineering (RE), LLMs have been applied to aid in the elicitation, documentation, and verification of software requirements. The ability of these models to understand complex specifications and provide coherent summaries or improvements has been highlighted in various studies. Cámara et al. proposed a conceptual framework to pave the way for standardization of the benchmarking process, ensuring consistent and objective evaluation of LLMs in software modeling~\citep{camara2024towards}. Ruan et al. presented an automated framework for requirement model generation that incorporates GPT-based zero-shot learning to extract requirement models from requirement texts and subsequently compose them using predefined rules~\citep{ruan2023requirements}. Arora et al. explored the use of LLMs in driving RE processes, focusing on the potential for requirements elicitation, analysis, specification, and validation, and found that LLMs can enhance several RE tasks by automating, streamlining, and augmenting human capabilities~\citep{arora2023advancing}. In~\citep{jin2024evaluation}, LLMs were used to extract problem diagrams from natural language documents to facilitate requirements extraction and modeling for cyber-physical systems. Tabassum et al. investigated the capabilities of OpenAI's GPT-4 and Google's Gemini for generating standard and UCM4IoT textual use cases by carrying out a comparative study using four IoT applications~\citep{tabassum2024using}. Chen et al. proposed to create class diagrams using LLMs with two sequential steps, including class generation and relationship generation, and the results show that their proposed approach outperforms the single-prompt-based approach by improving recall values and F1 scores in most systems for modeling the classes, attributes, and relationships~\citep{chen2024model}. Chen et al. conducted a comprehensive comparative study on automated domain modeling with LLMs (i.e., GPT-3.5 and GPT-4), and they found that LLMs demonstrate impressive capabilities on domain understanding, but are still impractical for fully automated domain modeling~\citep{chen2023automated}. Chaaben et al. evaluated the usefulness of a novel approach utilizing LLMs and few-shot prompt learning to assist in domain modeling, and the results show its usability and effectiveness~\citep{chaaben2024utility}. Netz et al. developed MAGDA, a user-friendly tool, through which they conduct a user study and assess the real-world applicability of our approach in the context of domain modeling, offering valuable insights into its usability and effectiveness~\citep{netz2024using}. For architecture design, Ahmad et al. conducted a preliminary study on using GPT to support a human-bot collaborative software architecting process~\citep{ahmad2023towards}. 

Several researchers have exploited the use of LLMs for education. Cipriano et al. experimented with three prominent LLMs to solve real-world object-oriented programming (OOP) exercises used in educational settings, subsequently validating their solutions using an Automatic Assessment Tool (AAT) and found that while the models frequently achieved mostly working solutions to the exercises, they often overlooked the best practices of OOP~\citep{cipriano2024llms}. In particular, our previous work explored how LLMs help undergraduate students create three types of UML diagrams, i.e., use case diagrams, class diagrams, and sequence diagrams, and synthesized the potential factors that affect their performance~\citep{Wang2024how}.

However, the application of LLMs specifically in evaluating and scoring software models, such as those represented in the Unified Modeling Language (UML), has not received comparable attention. Ferrari et al. investigated the capability of GPT to generate a specific type of diagram, i.e., UML sequence diagrams, from NL requirements by conducting a qualitative study in which they examine the sequence diagrams generated by GPT for 28 requirements documents of various types and from different domains~\citep{ferrari2024model}. Nevertheless, the limitations of this paper include a small sample size (only 28 requirement documents) and the subjective nature of manual evaluations, which may introduce bias. Additionally, the study only assessed UML sequence diagrams, leaving the generalizability to other modeling tasks in question. To evaluate the capability of LLM agents in correctly generating UML class diagrams, De Bari et al. collected 20 exercises from a diverse set of web sources and compared the UML class diagrams generated by a human and an LLM solver in terms of syntactic, semantic, pragmatic correctness, and the distance from a provided reference solution. They also suggested that future research should focus on evaluating the ability of LLMs to generate natural language modeling tasks and assess students' answers~\citep{de2024evaluating}. Our study is a direct implementation of that future direction, focusing on LLMs' performance in scoring UML class diagrams created by students. \textcolor{black}{Al-Ahmad et al. conducted a comprehensive empirical investigation into the effectiveness of GPT-4-turbo in generating four fundamental UML diagram types: Class, Deployment, Use Case, and Sequence diagrams \citep{al2025student}. This work provided empirically-grounded insights into the current capabilities and limitations of LLMs in software modeling education, but it concentrated on LLM-aided UML modeling, rather evaluating the created UML diagrams.}

Although the potential benefits of using LLMs to evaluate the quality and completeness of UML diagrams are clear, there is a significant gap in the existing literature that addresses this application. This highlights the need for further exploration of the methods that could take advantage of LLMs to enhance the evaluation process of UML diagrams, especially for the education and training in SE.

\section{Research Design}
\label{chap:case}
%This section describes the research objective, research data, experiment design, and evaluation criteria for the created UML models. 

%\subsection{Research Objective}
The main research objective of this work is to explore the feasibility and capability of using GPT to assess three types of UML diagrams, and to examine the implications of this approach for UML modeling education at the university level. For this purpose, we designed and conducted a series of experiments, as detailed in the following subsections.  

%\subsection{Experiment Design}
%To explore the capability of GPT in the assessment of UML models, we designed a 4-phase assessment framework consisting of a series of experiments, as shown in Figure 1. 

%\subsubsection{Phase I: Setting up UML modeling tasks}
\subsection{Research Data}
\label{sec:data}
The research data for this work is UML diagrams created by the students for their course reports. To standardize the assessment process and results, we design and perform a specified UML modeling task to get the UML diagrams to be graded. In this task, participants are required to create three types of UML diagrams, i.e., a use case diagram, a class diagram, and a sequence diagram, for the given case study titled \textit{Order Processing System} (see the description of this case study in~\cite{replpack}). 

The participants in this study are undergraduate students majoring in Software Engineering (SE) from Wuhan University, China. Before participating in this task, they had undergone a three-month training program in their \textit{Requirements Analysis and Modeling} course. During this course, they not only acquired knowledge of UML modeling but also honed their skills in crafting UML use case diagrams, class diagrams, and sequence diagrams for a designated case study. After receiving the UML modeling task mentioned above, the students have one week to complete and submit their project reports, including their UML diagrams and details of how to create these diagrams. As addressed in our previous work \citep{Wang2024how}, the students are allowed to create UML diagrams with the assistant of LLMs. This means that the UML diagrams in the project reports could be generated directly by LLMs, initially created by the students themselves and then improved by LLMs, or initially sketched by LLMs and then refined by the students. Note that these differences in the creation of UML diagrams in our research dataset are omitted in this work, because the objective of this exploratory study is to grade the created UML diagrams, rather than their constructing process.

The UML modeling task of \textit{Order Process System} was assigned to two groups of 30 students in 2023 and 2024 respectively, because the \textit{Requirements Analysis and Modeling} course is scheduled for the fifth semester of their undergraduate program. We initially collected 60 project reports. After a careful review, 20 reports were excluded because the UML diagrams were hand drawn, making it difficult for GPT to understand and analyze in our pilot study. Finally, UML diagrams \textcolor{black}{in 40 students' project reports} (20 students in each group) were selected as research data for this work. They were first graded by GPT and then manually evaluated by human experts. \textcolor{black}{To facilitate the evaluation conducted by both human experts and GPT,} these 120 UML diagrams ($3$ types $\times$ $20$ students $\times$ $2$ groups) \textcolor{black}{extracted from 40 project reports are stored in \texttt{.png} format and grouped into 40 folders (e.g., one folder per student) in the dataset. Note that in this grading task, both human experts and GPT are required to evaluate the images of UML diagrams rather than their textual descriptions, due to the following two reasons. First, the description of the case study provided in \citep{replpack} explicitly asks students to `draw' UML diagrams, which are the specified outcomes of UML modeling. Therefore, these images are what the educators need to grade during the course. Second, although GPT might perform better on textual descriptions of images, the transformation from images to textual descriptions might inherently result in information loss.}

%\subsubsection{Phase II: Defining evaluation criteria for UML models}
\subsection{Evaluation Criteria for UML Diagrams}
\label{sec:criteria}
Normally, the students' course assignments are graded according to suggested answers. Based on the suggested UML diagrams in \citep{replpack}, we define detailed evaluation criteria to grade the three types of UML diagram created by the students, as shown in Table~\ref{tab:uccriteria}, Table~\ref{tab:classcriteria}, and Table~\ref{tab:seqcriteria} respectively. To conduct a quantitative evaluation of UML diagrams, in the \textit{Grading} column of each table, \textit{Suggested answer} enumerates the essential elements that must be covered when evaluating each type of UML diagram. Correspondingly, \textit{Scores} specifies the points-awarding and points-deduction rules for applying each evaluation criterion. Both GPT-based and human-based assessment of UML diagrams should follow these eleven evaluation criteria. In terms of the number of evaluation criteria listed in these three table, the full scores of UML use case diagram, class diagram, and sequence diagram to be evaluated in this paper are 4 points, 4 points, and 3 points, respectively. Accordingly, the full score of each student report is 11 points. Note that in this paper, both the case study and its suggested UML diagrams (available at \citep{replpack}) have been provided in the supplementary teaching material of the \textit{Requirements Analysis and Modeling} course. In addition, the grading details in the following three tables are proposed by the first author and then refined during the discussion with the last two authors, according to their experience on grading this UML modeling task.

\begin{table}[!ht]
\scriptsize
\caption{Evaluation criterion for the created use case diagram}
\label{tab:uccriteria}
\centering
\begin{tabular}
{lp{2cm}p{3cm}p{8.5cm}}
\hline
No. & \textbf{Criteria} &  \textbf{Description} &\textbf{Grading} \\ \hline
\hline
UC1 & Complete identification of actors & Primary actors should exist and be named appropriately, aligning with the textual description.  & 
\textbf{Suggested answer}: Customer, Gold Customer, Accounting System \newline
\textbf{Scores}: The full score is 1 point, only if all the suggested three core actors are included. Deduct 0.2 points for missing each of the three suggested actors. Deduct 0.2 points for each of the additional and unreasonable actors. No points are deducted if the additional actors included in the use case diagram are consistent with the textual description of the case study. No points are deducted if the names of actors vary from the suggested answer but refer to the same roles.
%Different expressions with similar meanings are acceptable. 
 \\ \hline
UC2 & Complete identification of use cases & The use cases should be reflected in the text and represent \textcolor{black}{essential} functional requirements of the system.
   & \textbf{Suggested answer}: Place Order, Check Order Status, Cancel Order, Signup for Notification, Request Catalog, Return Product. Note: \textit{Modify Order} can replace the combination of \textit{Check Order Status} and \textit{Cancel Order}. \newline
\textbf{Scores}: The full score is 1 point, only if all the suggested use cases are included. Deduct 0.2 points for missing \textcolor{black}{each} of the following use cases: \textit{Place Order}, \textit{Check Order Status}, \textit{Cancel Order} (or \textit{Modify Order}), and \textit{Request Catalog}. Deduct 0.1 points for missing \textcolor{black}{each} of the following use cases: \textit{Signup for Notification} and \textit{Return Product}. Deduct 0.2 points for other unreasonable use cases. No points are deducted if the additional use cases are \textcolor{black}{not only reasonable but} consistent with the textual description of the case study. \textcolor{black}{No points are deducted if the names of use cases vary from suggested answer but refer to the same functionalities.}
  \\ \hline
UC3 & Accurate recognition of relationships among actors & The relationship between suggested actors should be identified correctly.
   & \textbf{Suggested answer}: The generalization relationship between \textit{Gold Customer} and \textit{Customer} should be accurately represented.\newline
  \textbf{Scores}: The full score is 1 point. Deduct 1 point for missing the generalization relationship between \textit{Gold Customer} and \textit{Customer}. \\ \hline
UC4 & Accurate recognition of relationships between actors and use cases & Suggested actors and use cases are correctly linked, aligning with the texts of the given case study. & \textbf{Suggested answer}: Based on UC1 and UC2, the identified use cases must be connected to the corresponding identified actors.\newline
  \textbf{Scores}: The full score is 1 point. Deduct 0.1 points for missing one necessary relationship between one actor and one use case. Deduct 0.1 points for introducing unnecessary and unreasonable relationships.
  \\ \hline
\end{tabular}%
\end{table}

%Especially, the suggested UML diagrams have been improved by the teachers in the curriculum instruction team of this course, according to their professional knowledge and teaching experience.} 

\begin{table}[htbp]
\scriptsize
\caption{Evaluation criterion for the created class diagram}
\label{tab:classcriteria}
\centering
\begin{tabular}{lp{2cm}p{3cm}p{9cm}}
\hline
No. & \textbf{Criteria} &  \textbf{Description} &\textbf{Grading} \\ \hline
\hline
CC1 & Complete identification of classes & Essential concepts in the case study should be covered. The classes are named clearly, descriptively, and meaningfully. & 
\textbf{Suggested answer}: Customer, GoldCustomer, CustomerRep, BackOrderNotifier, ShippingInfo, Order, Invoice, AccountingManager, Product, OrderProcessingManager \newline
\textcolor{black}{\textbf{Scores}: The full score is 1 point only if the diagram includes at least eight out of the aforementioned ten suggested classes. Deduct 0.2 points for missing each essential class if the number of included suggested classes is fewer than eight. Deduct 0.2 points for each unreasonable class that is clearly unrelated to the essential concepts in the case study. No points are deducted if the names of classes vary from the suggested answers but refer to the same essential concept. No points are deducted if the additional class is a reasonable refinement or supplement to the textual description of the case study.} 
 \\ \hline
CC2 & Appropriate attributes & There are necessary and reasonable attributes in the classes. The names of the attributes should be in accordance with the class.
   & \textbf{Suggested answer}: N/A \newline
\textbf{Scores}: The full score is 1 point. Each class should include at least one or two essential attributes. Based on CC1, if more than 70\% of the identified essential classes have one or two essential attributes, no points are deducted; otherwise, deduct a total of 0.2 points for the missing attributes.
For each unreasonable attribute that clearly does not belong to the responsibility of the class or is unrelated to the case study, deduct 0.1 points if 1-2 such attributes exist and 0.2 points if more than two exist. Deduct 0.1 points if the names of attributes are generally vague and confusing.
%\textbf{Scores}: The full score is 1 point, based on the rationality of the attributes defined in the class. Deduct 0.2 points for other unreasonable attributes. Deduct 0.1 points for too few attributes. 
  \\ \hline
CC3 & Appropriate operations & There are necessary and reasonable operations in the classes. The included operations should be named appropriately to be in accordance with the class and the object-oriented design principles.  
   & \textbf{Suggested answer}: N/A \newline
  \textbf{Scores}: The full score is 1 point.
  \textcolor{black}{ Each class with a defined responsibility should include at least one or two essential operations. \textcolor{black}{Based on CC1, if more than 70\% of the identified essential classes have one or two essential operations, no points are deducted; otherwise, deduct a total of 0.2 points for the missing operations.} 
  %If most essential classes in the diagram have seldom operations, deduct a total of 0.2 points. 
  For each unreasonable operation that clearly does not belong to the responsibility of the class or is unrelated to the case study, deduct 0.1 points if 1-2 such operations exist and 0.2 points if more than two exist. Deduct 0.1 points if the names of operations are generally vague and confusing.}
  %based on the rationality of the operations defined in the class. Deduct 0.2 points for other unreasonable operations. Deduct 0.1 points for too few operations. 
  \\ \hline
CC4 & Accurate recognition of relationships between classes & \textcolor{black}{For the essential classes identified by CC1, }the relationships between them are represented correctly and completely,
including generalization, association, aggregation, composition, etc. There are no obvious errors in the relationships. & \textbf{Suggested answer}: \textcolor{black}{GoldCustomer is the generalization of Customer, Order is the aggregation of Product, and Order is the aggregation of Invoice.} \newline
%Based on CC1, the identified classes must be connected with the appropriate relationships, using correct UML notations.\newline
  \textbf{Scores}: \textcolor{black}{The full score is 1 point. Deduct 0.2 points for missing each of the suggested relationships. Deduct 0.1 points for each relationship whose type is not appropriate. Deduct 0.2 points for each incorrect or unreasonable relationship in the diagram. No points are conducted if multiplicity is missed. No points are deducted for other reasonable relationship beyond the three essential ones.}
  %The full score is 1 point only if the relationships between the identified classes are reasonable and are represented completely with names and multiplicities. Deduct 0.2 points for other unreasonable relationships. Deduct 0.1 points for too few relationships. Deduct 0.1 points for missing multiplicities. Deduct 0.1 points for unreasonable names of relationships. 
  %This is evaluated by the modeling experts.
  \\ \hline
\end{tabular}%
\end{table}

\begin{table}[htbp]
\scriptsize
\caption{Evaluation criterion for the created sequence diagram}
\label{tab:seqcriteria}
\centering
\begin{tabular}{lp{2cm}p{3cm}p{9cm}}
\hline
No. & \textbf{Criteria} &  \textbf{Description} &\textbf{Grading} \\ \hline
\hline
SC1 & Correct identification of objects & Primary objects with appropriate names should be included, aligned with the textual description.  & 
\textbf{Suggested answer}: Customer, PlaceOrderScreen, OrderProcessingManager, InventoryManager, AccountingManager \newline
\textbf{Scores}: The full score is 1 point, only if \textcolor{black}{at least} four out of the suggested five objects are included. \textcolor{black}{Deduct 0.2 points for missing each essential object if the total number of included suggested classes is fewer than four.} No points are deducted if the additional objects included in the sequence diagram are consistent with the textual description of the case study.
 \\ \hline
SC2 & Correct identification of messages & The messages should be identified and named appropriately to represent the message type and purpose accurately and descriptively. The identified message should align with the textual description.
   & \textbf{Suggested answer}: N/A. \newline
\textbf{Scores}: The full score is 1 point, based on the rationality of the messages defined in the diagram. \textcolor{black}{Deduct 0.1 points for each missing message between the essential object identified based on SC1.} \textcolor{black}{Deduct 0.05 points for each message whose name is ambiguous or fails to accurately reflect its responsibility, with a total deduction not exceeding 0.3 points. Deduct 0.05 points for each message in which the parameters are unreasonable or obviously missed based on the case study, with a total deduction not exceeding 0.2 points. No points are deducted for either synchronous or asynchronous messages. Reusing the same message should not result in a deduction if it is a reasonable multiple call in the case study. The message constitutes unnecessary redundancy or logical errors is treated as an unreasonable message.}
%Deduct at most 0.4 points for unreasonable messages. 
  \\ \hline
SC3 & Correct identification of the order of identified messages. & The identified messages should be invoked in a specified order, according to the system process described in the texts of the case study. & \textbf{Suggested answer}: Based on SC1 and SC2, the order of messages is reasonable, aligning with the texts of the given case study. \newline
  \textbf{Scores}: The full score is 1 point, based on the rationality of the messages defined in the diagram. \textcolor{black}{If the overall sequence of the messages significantly deviates from the ones described in the case study, the total deduction should not exceed 0.4 points. Deduct 0.1 and 0.2 points for each specified reversals and illogical sequences respectively, with the total deduction not exceeding 0.4 points. Control flow of messages is not considered here.}
  \\ \hline
\end{tabular}%
\end{table}

Table~\ref{tab:uccriteria} lists four criteria to evaluate the created UML use case diagrams. More specifically, UC1 and UC2 are provided to evaluate the completeness of two essential use case modeling elements, i.e., actors and use cases, with suggested answers. UC3 and UC4 are designed to assess the rationality of the created diagram, based on the actors and use cases already identified. Therefore, the suggested answers to these two criteria are more subjective, compared to those for UC1 and UC2. 

The authors define four CCs to assess the created UML class diagrams, as shown in Table~\ref{tab:classcriteria}. CC1 is proposed to evaluate the classes, which are conceptual classes without attributes and/or operations. \textcolor{black}{CC2 and CC3 are proposed to evaluate the reasonability and sufficiency of class attributes and operations respectively, based on CC1. When human educators assess UML class diagrams, their judgments about whether the class attributes and operations are appropriate and sufficient rely on their professional expertise and experience, by considering both the identified classes and the overall coherence of the class diagram. Due to the complexity and subjectivity in specifying classes, attributes and operations, it is impractical to provide suggested answers to GPT. To balance the scoring consistency with contextual flexibility, the descriptions of CC2 and CC3 use the term `necessary' to enable both  human experts and GPT to accommodate diverse design choices while maintaining a shared standard of adequacy.} Note that in the grading details of CC2 and CC3, the 70\% proportion is specified based on the reference answer of the case study.
CC4 is designed to assess the rationality of the created diagram, based on the classes already identified. Therefore, the suggested answer to CC4 is more subjective, compared to those for CC1. \textcolor{black}{Note that multiplicities will not be evaluated in this task since they are ignored by either the suggested class diagram or most of the students reports, according to our pilot investigation.}

Table~\ref{tab:seqcriteria} shows the three SCs for grading the created UML sequence diagrams. SC1 is designed to evaluate whether enough essential objects are identified. In UML modeling, the operations of classes can normally be used to guide the identification of messages between objects. Therefore, there are no suggested answers to SC2, which is similar to that of CC3. SC3 is provided to assess the order of the messages identified by SC2 between the objects captured by SC1. As a result, the suggested answers are more subjective than those of SC1. \textcolor{black}{Note that according to the suggested answer, the complex modeling elements in UML sequence diagrams, such as control flow and `ref' frame, will not be evaluated in this task, since they are ignored by either the suggested sequence diagram or most of the students reports, according to our pilot investigation.}

%\subsubsection{Phase III: Assessing UML models with GPT}
\subsection{GPT-based Assessment}
\label{gptprompt}
In our work, GPT-4o \textcolor{black}{is the version of GPT} that was used to evaluate the three types of UML diagrams created by the 40 students. The reason is that in our pilot study, GPT-4o outperformed the other LLMs when evaluating UML diagrams. \textcolor{black}{Note that in this paper, all subsequent references to GPT refer specifically to the GPT-4o model.} The prompt for the GPT assessment is detailed in the gray box below. \textcolor{black}{To improve GPT's performance,} the prompt is designed to integrate the role that GPT would like to play in this task, the description and the suggested answer of the UML modeling task, the grading criteria defined in Section~\ref{sec:criteria}, etc. \textcolor{black}{The rationale of the prompt design is to to follow role-play prompting, acting as a more effective trigger for the Chain-of-Thought process \citep{kong2024}.} Besides these pieces of knowledge required to evaluate the UML diagrams created by the students, the prompt specifies the steps followed by the teachers to manually grade these UML diagrams. This helps GPT to work as educators \textcolor{black}{and perform better} in assessing UML diagrams.

\begin{tcolorbox}[breakable]
%\small
\textbf{Prompt for assessing UML diagrams:}

I am a computer science lecturer at a university, teaching a course called \textit{Requirements Analysis and Modeling}. I made an assignment on the creation of UML diagrams. Now I need some help in grading these assignments submitted by the students.\\
\\
The assignment is provided as follows: There is a textual description of a case study, i.e., an Order Processing System, followed by three questions on designing a use case diagram, a class diagram, and a sequence diagram for this case study. Firstly, I will send you the textual description and three questions of this assignment, followed by the reference solutions. Then, the evaluation criteria for grading these UML diagrams will be provided. Finally, I will give you the UML diagrams created by the students, and you need to grade them reasonably based on your knowledge of UML diagrams and the following information I provide.\\
\\
Here is the problem statement:

Case Study: Order Processing System

\textit{Consider the following problem description: A mail-order company wants to automate its order processing. The initial version of the order processing system should be accessible to customers via the web. Customers can also call the company by phone and interact with the system via a customer representative. It is highly likely that the company will enhance this system in upcoming years with new features.}

\textit{The system allows customers to place orders, check the status of their orders, cancel an existing order, and request a catalog. Customers may also return a product, but this is only possible through the phone, not available on the web. When placing an order, the customer identifies himself either by means of a customer number (for existing registered customers) or by providing his name and address. He then selects a number of products by giving the product number or by selecting products from the online catalog. For each product, information such as price, a description, and a picture (only on demand as they are usually high-resolution images of large size) is presented to the customer. Also, the availability of the product is obtained from the inventory. The customer indicates whether he wants to buy the product and in what quantity. When all desired products have been selected, the customer provides a shipping address and a credit card number and a billing address (if different from the shipping address). Then an overview of the ordered products and the total cost are presented. If the customer approves, the order is submitted. Credit card number, billing address, and a specification of the cost of the order are used on the invoice, which is forwarded to the accounting system (an existing software module). Orders are forwarded to the shipping company, where they are filled and shipped.}

\textit{Customers who spent over a certain amount within the past year are promoted to be gold customers. Gold customers have additional rights such as being able to return products in an extended time period as well as earning more bonus points with each purchase. In addition, in cases where a product is on back order, gold customers have the option to sign up for an email notification for when the particular product becomes available.}

\textit{(1) Identify actors and use cases for the system described above and show them on a UML Use Case Diagram.}

\textit{(2) Perform a quick application domain analysis to come up with an object model for the above system. Express your findings with a UML Class Diagram, making sure to identify any critical operations of classes.}

\textit{(3) Consider the following use case scenario (for the use case `place order':}

\textit{Ali is an existing customer of the order processing company described earlier, registered with their website. Also assume that having browsed the printed catalog he has, he has already identified the two items (including their prices) he likes to buy from the company's website using their product numbers (i.e., 2 and 9). First, he tries to buy one of product 2, but it is listed as unavailable in the inventory. Then, he adds two quantities of product 9, which turns out to be available, to his basket. He is then asked to confirm his registered shipping and billing addresses and credit card information from the customer database. He completes the order by clicking the Submit button. You may ignore the processing of customer authentication.}

\textit{Draw a UML Sequence Diagram for this particular scenario. You may use any software/solution domain objects if needed as well.}\\
\\
Please read these texts first. Then, I will send you the reference answers to these three questions.

\textcolor{black}{[The images of the three reference UML diagrams will be provided here.]}

The 3 images are the reference answers for Questions (1), (2), and (3). Please read them. Next, I will send you the evaluation aspects and scoring criteria for each question.\\
\\
For Question (1) - Use Case Modeling, the total score is 4 points, based on the evaluation criteria UC1, UC2, UC3, and UC4.

Grading based on UC1: \textcolor{black}{[The text in Line 1, Column 4 of Table~\ref{tab:uccriteria} will be listed here.]}

Grading based on UC2: \textcolor{black}{[The text in Line 2, Column 4 of Table~\ref{tab:uccriteria} will be listed here.]}

Grading based on UC3: \textcolor{black}{[The text in Line 3, Column 4 of Table~\ref{tab:uccriteria} will be listed here.]}

Grading based on UC4: \textcolor{black}{[The text in Line 4, Column 4 of Table~\ref{tab:uccriteria} will be listed here.]}\\
\\
For Question (2) - Class Diagram modeling, the total score is 4 points, based on the evaluation criteria CC1, CC2, CC3, and CC4.

Grading based on CC1: \textcolor{black}{[The text in Line 1, Column 4 of Table~\ref{tab:classcriteria} will be listed here.]}

Grading based on CC2: \textcolor{black}{[The text in Line 2, Column 4 of Table~\ref{tab:classcriteria} will be listed here.]}

Grading based on CC3: \textcolor{black}{[The text in Line 3, Column 4 of Table~\ref{tab:classcriteria} will be listed here.]}

Grading based on CC4: \textcolor{black}{[The text in Line 4, Column 4 of Table~\ref{tab:classcriteria} will be listed here.]}\\
\\
For Question (3) - Sequence Diagram modeling, the total score is 3 points, based on the evaluation criteria SC1, SC2, and SC3.

Grading based on SC1: \textcolor{black}{[The text in Line 1, Column 4 of Table~\ref{tab:seqcriteria} will be listed here.]}

Grading based on SC2: \textcolor{black}{[The text in Line 2, Column 4 of Table~\ref{tab:seqcriteria} will be listed here.]}

Grading based on SC3: \textcolor{black}{[The text in Line 3, Column 4 of Table~\ref{tab:seqcriteria} will be listed here.]}\\
\\
It is important to emphasize that when grading, please do not be overly rigid. Modeling elements in the UML diagrams created by the students do not need to exactly match the suggested answers in terms of naming, as long as the meaning and design are consistent with the requirements.

Here are the three UML diagrams to be graded:

\textcolor{black}{[The images of the three UML diagrams created by one student will be provided here.]}
\end{tcolorbox}

In each round of GPT-based assessment, only the three UML diagrams created by one student are graded. That is, a 40-round evaluation should be performed to cover the research dataset~\citep{replpack}. \textcolor{black}{The operational parameters used in this GPT-based assessment are listed in Table~\ref{tab:para}.} 

\begin{table}[htbp]
\scriptsize
\centering
\caption{\textcolor{black}{Operational parameters used in GPT-based assessment}}
    \label{tab:para}
\begin{tabular}{p{5cm}p{3cm}}
        \hline
        \textbf{Parameter} & \textbf{Value} \\ \hline
        Number of runs per time & 1 \\ \hline
        Result aggregation policy& N/A \\ \hline
        Use of a seed & No\\ \hline
        Temperature & 0.1 \\ \hline
        Top-p & N/A \\ \hline
        Time window of data collection & 2023/05/01-2023/06/30, 2024/05/01-2024/06/30 \\ \hline
\end{tabular}
\end{table}

To facilitate the GPT-based assessment, we designed and implemented a Python program for the semi-automated assessment of UML diagrams in the dataset. The inputs for this program are as follows.

\begin{mdframed}
    \textcolor{black}{\textbf{Inputs:}}
    \begin{itemize}
        \item \textbf{student\_id}, referring to the name of the folder containing three UML diagrams created by this student
        \item \textbf{case\_study.md} referring to the description of \textit{Order Processing System} in markdown format
        \item \textbf{criteria.md}, referring to evaluation criteria defined in Table~\ref{tab:uccriteria}, Table~\ref{tab:classcriteria}, and Table~\ref{tab:seqcriteria} in markdown format
        \item \textbf{Student's UML diagrams}, referring to images of the three UML diagrams contained in the folder titled with \textbf{student\_id}, imported by the code snippet below \\
        {\small
        \texttt{
        image\_labels = ["Student Use Case Diagram", "Student Class Diagram", \\"Student Sequence Diagram"]\\
        for i, (label, image\_b64) in enumerate(zip(image\_labels, student\_images\_base64)):\\
           \hspace*{4em} user\_content.extend([\\
           %    \hspace*{4em}\hspace*{4em} \{"type":"text", "text":f"\n**\{label\}:**"\},\\
               \hspace*{4em}\hspace*{4em} \{"type":"image\_url",\\
                   \hspace*{4em}\hspace*{4em} "image\_url": \{"url": 
                   f"data:image/png;base64,\{image\_b64\}"\}])
            }
        }
        
        \item \textbf{Suggested answer}, imported by the code snippet below\\
        {\small
        \texttt{
        std\_images\_dir = self.settings.get\_std\_images\_path()\\
            \hspace*{4em} ref\_use\_case = std\_images\_dir/`u.png'\\
            \hspace*{4em} ref\_class = std\_images\_dir/`c.png'\\
            \hspace*{4em} ref\_sequence = std\_images\_dir/`s.png'}
        }
    \end{itemize} 
\end{mdframed}

\textcolor{black}{The output of our GPT-based assessment is a JSON file, as shown below. This JSON file lists the reasons and the deducted points after applying each evaluation criterion in Table~\ref{tab:uccriteria}, Table~\ref{tab:classcriteria}, and Table~\ref{tab:seqcriteria}. Appendix A of this paper provides an example of the output of GPT-based assessment of the three UML diagrams.}

\begin{mdframed}
%\textcolor{black}{\textbf{Output:}}\\
    {\small
        \texttt{
  "student\_id": "{student\_id}",\\
   \hspace*{4em}"uc1\_deductions": [\{"reason":"", "points":\},\\
    \hspace*{13em} \{{"reason":"", "points": \}}],\\
\hspace*{4em}"uc2\_deductions":[\{"reason":"", "points":\}],\\
\hspace*{4em}"uc3\_deductions":[\{"reason":"", "points":\}],\\
\hspace*{4em}"uc4\_deductions":[\{"reason":"", "points":\}],\\
\hspace*{4em}"cc1\_deductions":[\{"reason":"", "points":\}],\\
\hspace*{4em}"cc2\_deductions":[\{"reason":"", "points":\}],\\
\hspace*{4em}"cc3\_deductions":[\{"reason":"", "points":\}],\\
\hspace*{4em}"cc4\_deductions":[\{"reason":"", "points":\}],\\
\hspace*{4em}"sc1\_deductions":[\{"reason":"", "points":\}],\\
\hspace*{4em}"sc2\_deductions":[\{"reason":"", "points":\}],\\
\hspace*{4em}"sc3\_deductions":[\{"reason":"", "points":\}],
        }}
\end{mdframed}

\subsection{Human-based Evaluation}
To further evaluate the performance of GPT in grading UML use case diagrams, class diagrams, and sequence diagrams, we conduct a human-based assessment on the UML diagrams created by the students in Section~\ref{sec:data} and graded by GPT in Section~\ref{gptprompt}.

\textcolor{black}{The authors of this paper were engaged in grading of 120 UML diagrams manually. More specifically, the second author worked as the teaching assistant of \textit{Requirements Analysis and Modeling} course in 2024. The remaining three authors are affiliated with universities and possess a minimum of three years' teaching experience in this course.} First, the first two authors randomly selected four students' project reports and graded their UML diagrams independently, based on their understanding of the evaluation criteria defined in Section~\ref{sec:criteria}. Then, all four authors had a short meeting to resolve the unclear issues and inconsistent understanding when grading these diagrams. After achieving consistency in grading the three types of UML diagrams, the second and the fourth authors gave scores to all the UML diagrams in the project reports of all 40 students. Finally, the first author randomly reviewed the scores of four project reports to ensure consistency in the scores and grading details, according to the proposed evaluation criteria.   

\section{Results}
\label{sec:results}
In this section, the experimental results intend to answer the following two research questions (RQs). 

\textbf{RQ1:} \textit{Can GPT be used to assess UML diagrams?} This RQ is designed to investigate the practicality of GPT in assessing UML diagrams, compared with manually grading these UML diagrams by professionals and/or educators. 

\textbf{RQ2:} \textit{How competent is GPT in assessing UML diagrams?} This RQ is designed to evaluate the strengths and weaknesses of GPT in assessing different types of UML diagrams. 

\textcolor{black}{The answers to these two RQs are based on the comparison between the scores and the corresponding reasons given by human experts and GPT, by referring to the same suggested answers. Since the objective of this paper is to explore whether GPT can (partially) replace human experts in assessing UML diagrams, the scores and reasons provided by human experts are treated as the baseline when evaluating the performance of GPT in this grading task. A smaller evaluation difference between GPT and human experts suggests that GPT has a higher potential and is more feasible for (partially) replacing human educators in grading UML diagrams during regular teaching.}

\subsection{Answer to RQ1}\label{RQ1}
Table~\ref{tab:score} lists the scores of 40 project reports that GPT and the authors graded, respectively. We found that compared to the grades generated by GPT, human scorers always gave higher scores to each of the three UML diagrams. 
More specifically, the highest and the lowest scores of UML Use Case diagrams graded by GPT and human scorers are the same, but the average score given by human beings is \textcolor{black}{13.69\% higher} than GPT`s. As for UML Class diagrams, the lowest and average scores graded by the authors are \textcolor{black}{10\% and 12.94\% }higher than that generated by GPT, respectively. Differently, the highest, lowest and average scores of UML Sequence diagrams graded by the authors are all higher than those generated by GPT, \textcolor{black}{accounting for 25\%, 24.24\%, and 24.26\% respectively}. The score of the whole project report sums the scores of the included UML use case diagram, class diagram, and sequence diagram. As the last three columns of Table~\ref{tab:score} show, the highest, lowest, and average scores of the whole report graded by the authors are \textcolor{black}{4.6\%, 27.62\%, and 16.39\%} higher than those generated by GPT, respectively.

\begin{table}[h]
\scriptsize
\caption{\textcolor{black}{Scores of the created UML diagrams}}
\label{tab:score}
\centering
\begin{tabular}{p{1cm}p{0.5cm}p{0.5cm}p{0.5cm}p{0.5cm}p{0.5cm}p{0.5cm}p{0.5cm}p{0.5cm}p{0.5cm}p{0.5cm}p{0.5cm}p{0.5cm}}
\hline
\multirow{2}{*}{Scorer} & \multicolumn{3}{c}{\textbf{Use Case Diagram}} &  \multicolumn{3}{c}{\textbf{Class Diagram}} & \multicolumn{3}{c}{\textbf{Sequence Diagram}} & \multicolumn{3}{c}{\textbf{Score of Report}} \\ 
 & H & L & Avg.  & H & L & Avg.  & H & L & Avg.  & H & L & Avg.\\ \hline
\hline
Human Beings & 3.9 & 1.9 & 2.99 & 3.7 & 2.2 & 2.88 & 3 & 1.8 & 2.51 & 10.25 & 6.7 & 8.38
  \\ \hline
GPT & 3.9 & 1.3 & 2.63 & 3.7 & 2 & 2.55 & 2.4 & 1.45 & 2.02 & 9.8 & 5.25 & 7.20
 \\ \hline
Diff. & 0 & +0.6 & +0.36 & 0 & +0.2 & +0.33 & +0.6 & +0.35 & +0.49 & +0.45 & +1.45 & +1.18
  \\ \hline
\multicolumn{13}{l}{H: Highest score, L: Lowest score, Avg.: Average score}
\end{tabular}
\end{table}

\textcolor{black}{Then, we intend to explore the disparity in the scores given by the human scorer and GPT for each of the 40 student reports. In this paper, a positive score difference indicates that the score given by human experts surpasses the one provided by GPT, a negative score difference indicates that the score given by human experts is lower than the one provided by GPT, and a zero-difference means that human experts and GPT give the same score to a specific UML diagram. }

\begin{figure}[h]
\centering
\includegraphics[width=1\textwidth]{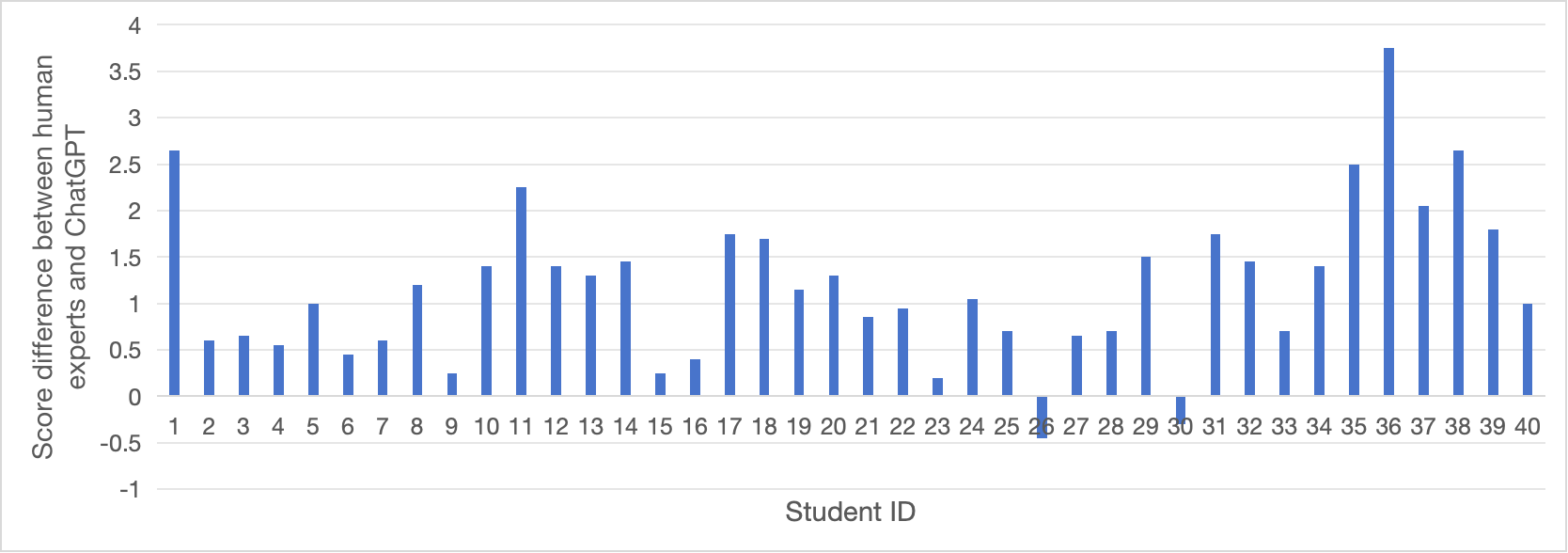}
\caption{\textcolor{black}{Score difference of the whole report between human experts and GPT over 40 students}}
\label{fig:scorediff1}
\end{figure}

\textcolor{black}{Figure~\ref{fig:scorediff1} shows the difference in the score of each student report that contains three UML diagrams. We observed that only two of the 40 student reports got higher scores from GPT. The maximum score difference between human experts and GPT is 3.75, and the positive score differences of 23 student reports are greater than 1 point. Whereas, the minimum score difference is 0.2.}

Table~\ref{tab:diffscore} zooms in on the disparity in the scores of three types of UML diagrams given by human experts and GPT. More specifically, we found that three UML use case diagrams, four class diagrams, and two sequence diagrams in 22.5\% of the students reports (9 of 40) were graded by both human experts and GPT with the same scores. Compared to UML use case diagrams and class diagrams, there are more positive and less negative score differences in grading UML sequence diagrams, accounting for 92.5\% and 2.5\% of 40 diagrams, respectively. 

\begin{table*}[h]
\scriptsize
    \caption{\textcolor{black}{Disparity in scoring three types of UML diagram}}
    \label{tab:diffscore}
    \centering
    \begin{tabular}{l c c c}
        \hline
        \textbf{Type of Score Difference }& \multicolumn{3}{c}{\textbf{Occurrence}}  \\ 
         & \textbf{Use Case Diagram} & \textbf{Class Diagram}  & \textbf{Sequence Diagram} \\ \hline
        \textbf{Zero-difference} & 3 & 4 & 2 \\ \hline
        \textbf{Positive} & 32 & 32 & 37  \\ \hline
        \textbf{Negative} & 5 & 4 & 1 \\ \hline
    \end{tabular}
\end{table*}

\begin{figure}[h]
\scriptsize
\centering
\includegraphics[width=1.0\textwidth]{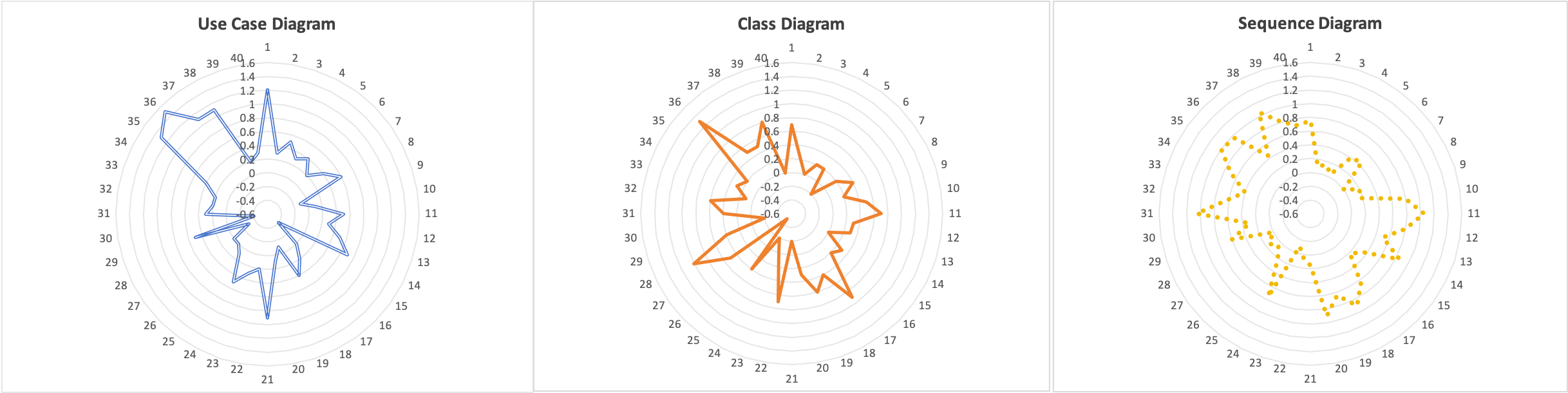}
\caption{\textcolor{black}{Distribution of score difference in three types of UML diagram created by 40 students}}
\label{fig:scorediff2}
\end{figure}

\textcolor{black}{Figure~\ref{fig:scorediff2} shows the distribution of score differences in three UML diagrams of 40 student reports. Compared to the grading of UML sequence diagrams, the fluctuations in the positive and negative score differences are more pronounced. This implies a significant divergence when human experts and GPT scored the same use case diagrams and class diagrams. Another observation is that the score discrepancies are greater when grading UML use case diagrams, in contrast to the relatively narrower score differences reported in the assessment of UML sequence diagrams.}

%\begin{center}
%\fcolorbox{black}{gray!10}{\parbox{0.97\linewidth}{\textbf{Key Findings of RQ1:} We found that based on the detailed evaluation criteria, GPT \textcolor{black}{is effective} in assessing UML use case diagram, class diagram, and sequence diagram. However, GPT usually generated lower scores than the human experts did, and even but the grading difference between GPT and human experts is not great. In addition, we identified three types of assessment discrepancies when GPT assesses UML models, compared to the assessment performed by human experts.
%\end{center}

\begin{center}
\fcolorbox{black}{gray!10}{\parbox{0.97\linewidth}{\textbf{Key Findings of RQ1:} We found that based on the detailed evaluation criteria, GPT \textcolor{black}{is effective} in assessing UML use case diagram, class diagram, and sequence diagram. \textcolor{black}{However, GPT usually generates lower scores than human experts, and the average grading difference between human experts and GPT is 1.18 points (full score is 11 points). In addition, the performance of GPT varies in evaluating different types of UML diagrams, according to the score discrepancy of each UML diagram in the 40 student reports.} }}
\end{center}

\subsection{Answer to RQ2} 
\label{RQ2}
\textcolor{black}{To further explore GPT's competency in understanding and applying the evaluation criteria defined in Section~\ref{sec:criteria} to grade UML diagrams, we performed an in-depth and insightful analysis of the score differences reported in Section~\ref{RQ1}. For this purpose, we manually reviewed the results of the GPT-based assessment, i.e., the JSON files that are exemplified in Section~\ref{gptprompt}, of 120 UML diagrams, especially the reasons that GPT adheres to the corresponding deducted points. Through comparative analysis with the reasons provided by human experts, we summarize the following five types of assessment discrepancies that cause differences in the scores generated by GPT.} 
\begin{itemize}
    \item \textcolor{black}{\textbf{Perfect Match.} It refers to the situation where the UML diagram whose grading results and reasons for point deduction provided by GPT are the same as those by human experts.}
    \item \textbf{Misunderstanding.} It indicates the case where GPT misunderstands the grading instruction of the evaluation criteria defined in Section~\ref{sec:criteria} or uses inappropriate evaluation criteria for the UML diagrams to be graded. 
    %Taking CC1 as an example, GPT sometimes deducts points when the total number of the identified classes is between 8 and 10. In this case, GPT misunderstood the implied meaning of CC1, i.e., ``\textit{Deduct 0.2 points for each missed class if the number of identified classes is less than eight}''. This implies that the class diagram containing 8 to 10 classes is reasonable.
     \item \textbf{Overstrictness.} It means that when grading UML diagrams, GPT is too strict by absolutely following the suggested answers, rather than allowing alternative answers with similar expressions.
     %One typical example is the enumerated values of the class `OrderStatus' in the suggested class diagram. Considering the case without specifying the enumerated values for `OrderStatus', GPT deducted points by strictly following the suggested answer, but human experts will not deduct points because they agree that enumerated values are not mandatory for class design.
     \item \textbf{Wrong Identification.} It reflects the errors made in identifying the modeling elements in the specified UML diagrams.
     %For example, GPT did not identify the generalization between two use cases, but this relationship was explicitly represented in the specified use case diagram.
     \item \textcolor{black}{\textbf{Reasonable Identification.} It refers to the situation in which GPT deducts points based on its precise understanding and proper utilization of the evaluation criteria. However, the modeling element that should lead to point deduction is not identified by human experts. }
     %However, the modeling element that should lead to point deduction is not identified by human experts. For example, GPT deducted 0.2 points for the missing essential actor \textit{Accounting System}, but this missing actor is not identified by human experts.}
\end{itemize}

%More specifically, Table~\ref{tab:discrepancies} shows varying performance that GPT exhibited when grading different types of UML models. We observed that `Overstrictness' is the most common type of evaluation discrepancy that GPT exposed when applying the proposed evaluation criteria in evaluating UML models. That is to say, GPT worked too strictly when assessing these three types of UML models, accounting for 13 use case diagrams, 12 class diagrams, and 12 sequence diagrams. As listed in the first row of Table~\ref{tab:discrepancies}, four use case diagrams, eight class diagrams, and six sequence diagrams were graded with different scores because GPT `misunderstood' the specified evaluation criteria it employed. In addition, some modeling elements in five use diagrams and two class diagrams were `wrongly identified' by GPT, as reported in the third row of Table~\ref{tab:discrepancies}. In the following subsections, we analyze the competency of GPT in assessing the UML use case diagram, class diagram, and sequence diagram respectively. (\textcolor{black}{To be reivsed if necessary})

Note that the analysis on the reasons for deducting points and the labeling of these five types was performed by the first author manually. Before that, a pilot analysis was executed by the first author and the third author to enhance the quality of these type labels. \textcolor{black}{For each of the 11 evaluation criteria, illustrative examples are provided in our replication package \citep{replpack}, including (1) the UML diagram being evaluated - assessed independently by both human experts and GPT, (2) the scores given by human experts and GPT, (3) the reasons underlying those scores, and (4) the type of discrepancy identified between the human and GPT evaluations. }

\begin{comment}

\begin{table*}[h]
\scriptsize
    \caption{Competency of GPT in assessing three types of UML Models}
%    \caption{Discrepancies in Evaluation of the UML Models}
    \label{tab:discrepancies}
    \centering
    \begin{tabular}{l c c c}
        \hline
        Type of Evaluation Discrepancies & \multicolumn{3}{c}{Occurence}  \\ 
         & Use Case Diagram & Class Diagram  & Sequence Diagram \\ \hline
        Misunderstanding & 4 & 8 & 6 \\ \hline
        Overstrictness & 13 & 12 & 12  \\ \hline
        Wrong Identification & 5 & 2 & 0  \\ \hline
%        Reasonable Disagreement & 5 & 8 & 7  \\ \hline
        Perfect Match & 21&19& 22  \\ \hline
    \end{tabular}
\end{table*}
\end{comment}

\subsubsection{Competency of GPT in assessing use case diagram}
\label{competency4UC}
This subsection analyzes the competency of GPT when it employs the four UCs defined in Table~\ref{tab:uccriteria} to grade \textcolor{black}{40} UML use case diagrams. In particular, five types of evaluation discrepancies identified in Section~\ref{RQ1} are refined to expose the weakness and strength of GPT in understanding and using which evaluation criteria. 
\begin{table*}[t]
\scriptsize
    \caption{\textcolor{black}{Occurrence of different types of discrepancies in the assessment of use case diagrams}}
    \label{tab:disuc}
    \centering   
    \begin{tabular}{r c c c c }
    \hline
    \textbf{Type of Evaluation Discrepancies (Occur.)} & \textbf{UC1} & \textbf{UC2} & \textbf{UC3} & \textbf{UC4} \\ \hline
    Perfect Match (68) & 17 & 10 & 37 & 4 \\ \hline
    Wrong Identification (36) & 3 & 7 & 2 & 24 \\ \hline
    Overstrictness (36) & 13 & 20 & 1 & 2 \\ \hline
    Misunderstanding (17) & 5 & 5 & 0 & 7 \\ \hline
    Reasonable Identification (22) & 3 & 6 & 0 & 13 \\ \hline
    \textit{Total:} & 41 & 48 & 40 & 50 \\ \hline
    \end{tabular}
\end{table*}

Table~\ref{tab:disuc} summarizes the occurrences of different types of evaluation discrepancies for each grading criterion in the assessment of \textcolor{black}{40} use case diagrams. When analyzing the reasons provided by GPT, we found that a single rationale for deducting points from a specific UML use case diagram, based on a particular evaluation criterion, can encompass multiple types of assessment discrepancies. More specifically, GPT performed completely the same as human experts when evaluating the actors in 17 use case diagrams with UC1. The second common assessment discrepancy reported in using UC1 is \textit{`Overstrictness'}. Regarding GPT's evaluation with UC2, \textit{Overstrictness} is the most frequent evaluation discrepancy, which occurs in half of the 40 UML use case diagrams. It is interesting to observe GPT's outstanding performance in using UC3, since \textit{`Perfect Match'} was reported in 37 out of the 40 UML use case diagrams. However, GPT often deducts points due to wrong identifications of the relationship between actors and use cases, which occurs in 24 out of the 40 use case diagrams.
Note that the subtotals of the occurrence of evaluation discrepancies using UC2 and UC4 are 48 and 50, respectively. This indicates that more than one type of evaluation discrepancies are more likely to occur when GPT applies UC2 and UC4 to evaluate use case diagrams created by the 40 students. 
%Note that the subtotal of occurrences with the evaluation discrepancy using UC1 is more than 40. This indicates that two types of evaluation discrepancies were matched when GPT assessed one use case diagram with UC1.  

\begin{figure}[h]
\scriptsize
\centering
\includegraphics[width=1.0\textwidth]{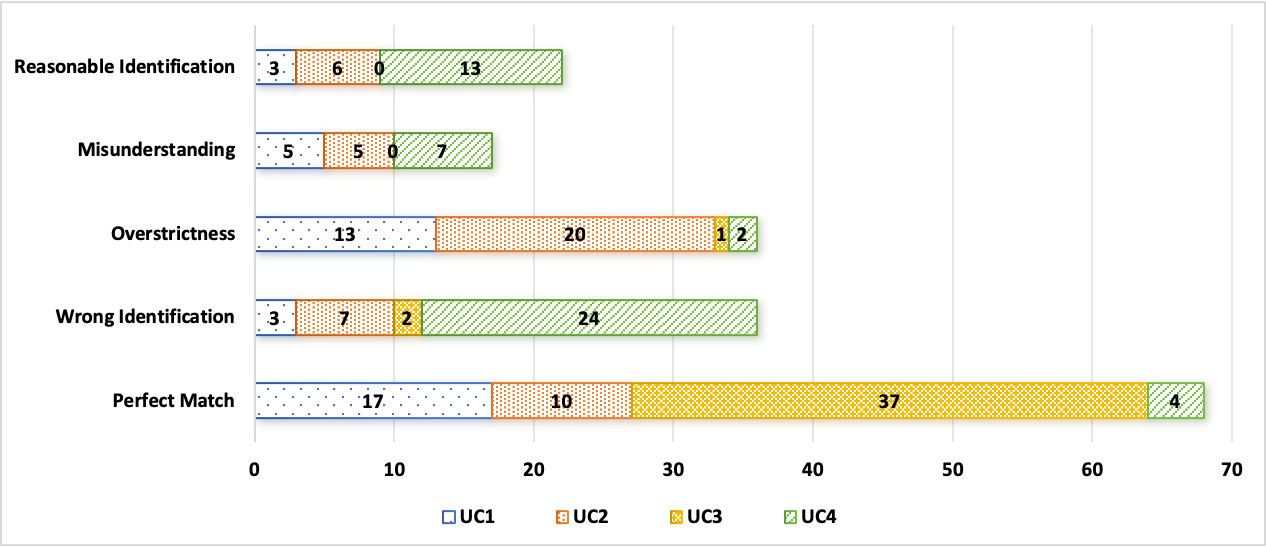}
\caption{\textcolor{black}{Occurrence distribution of the five discrepancy types in grading 40 UML use case diagrams with four UCs}}
\label{fig:ucsta2}
\end{figure}

\textcolor{black}{Figure~\ref{fig:ucsta2} zooms in on the distribution of occurrences of the five types of discrepancies when human experts and GPT independently use four UCs to grade 40 UML use case diagrams.} The observations and the corresponding examples are listed as follows.

\begin{itemize}
    \item \textbf{Perfect Match} 
   \textcolor{black}{is the most prominent type of discrepancy reported in the GPT-based assessment of UML use case diagrams, with 68 occurrences shown in Table~\ref{tab:disuc}. Especially, as the last bar of Figure~\ref{fig:ucsta2} presents, 54.4\% of \textit{`Perfect Match'} (37 of 68 occurrences) is reported in the use of UC3. This indicates that GPT performs well in recognizing relationships between actors (i.e., generalization between `Gold Customer' and `Customer' in our case study),} aligning closely with the grading criteria and human experts. \textcolor{black}{The second and third largest proportions of \textit{`Perfect Match'} fall into the use of UC1 (identifying actors) and UC2 (identifying use cases), accounting for 25\% and 14.7\% respectively. However, GPT's capability in understanding and using UC4 was the weakest.}

\begin{mdframed}
    \textbf{Example:} During the evaluation based on UC3, both human scorers and GPT can identify the missed relationship between `Gold Customer' and `Customer'. \textcolor{black}{In the evaluation based on UC1, both human experts and GPT can identified the missed essential actors, such as `Accounting System' or `Gold Customer'.}
\end{mdframed}

    \item \textbf{Wrong Identification}   
    \textcolor{black}{is the second most frequent type of evaluation discrepancy in the GPT-based assessment of UML use case diagrams, with 36 occurrences in Table~\ref{tab:disuc}. Specifically, the fourth bar in Figure~\ref{fig:ucsta2} shows two-thirds of \textit{`Wrong identification'} occurred in the evaluation based on UC4 (24 of 36 occurrences). This type of discrepancy reflects the main limitations of GPT in correctly identifying the relationships between specified actors and use cases. Another observation is that compared to UC1 and UC3, it is more common for GPT to misidentify} some essential use cases by failing to accurately map them to the given functional requirements. 

 \begin{mdframed}
     \textbf{Example:} \textcolor{black}{In the evaluation based on UC4, GPT deducts points for missing relationships between specified actors and use cases, but theses relationships existed in the UML use case diagram.} In the evaluation using UC3, the inheritance relationship between `Golden Customer' and `Customer' did exist in the UML use case diagram, but GPT failed to identify it and then deducted points.
 \end{mdframed}
    
    \item \textbf{Overstrictness} \textcolor{black}{is also the second frequent type of discrepancy found in the GPT-based assessment of UML use case diagrams, with 36 occurrences in Table~\ref{tab:disuc}. As the third bar of Figure~\ref{fig:ucsta2} shows, 55.6\% and 36.1\% of \textit{`Overstrictness'} respectively occurs in the evaluation based on UC2 and UC1, i.e., to identify the essential use cases and actors in the UML use case diagram.} 
    
\begin{mdframed}
   \textbf{Example:} Considering the evaluation based on UC1, some students named certain actors \textcolor{black}{in Chinese or with English} terms different from the ones in the reference answer. The human scorers treat the naming as reasonable and acceptable. However, the evaluation conducted by GPT was rigorous, resulting in points deduction. When evaluating the UML use case diagram where one use case is named `Check Order' instead of `Check Order Status', for example, GPT deducts points but human scorers do not.
\end{mdframed}

    \item \textbf{Misunderstanding} \textcolor{black}{is the discrepancy type with the lowest occurrence reported in the GPT-based assessment of UML use case diagrams, with 17 occurrences in Table~\ref{tab:disuc}. As the second bar of Figure~\ref{fig:ucsta2} shows, GPT misunderstood UC1 and UC2 in evaluating five use case diagrams, and misunderstood UC4 in grading seven use case diagrams.}

\begin{mdframed}
    \textbf{Example:} \textcolor{black}{Considering the use cases that are not only unreasonable but excluded in the suggested answer, UC2 is defined to deduct 0.2 points for all these use cases (pl.). However, GPT deducts 0.2 points for each of these use cases.}
\end{mdframed} 

    \item\textbf{\textcolor{black}{Reasonable Identification}}
    \textcolor{black}{occurs 22 times in evaluating UML use case diagrams, as shown in Table~\ref{tab:disuc}. The first bar of Figure~\ref{fig:ucsta2} represents that the \textit{`Reasonable Identification'}-typed discrepancy effectively pinpointed some actors, use cases, and the links between them that were overlooked by human experts but essential for inclusion in the use case diagrams. Specifically, GPT's understanding and use of UC1, UC2, and UC4 provided supplementary insights to the evaluation of three, six, and thirteen use case diagrams, respectively.}
    
\begin{mdframed}
    \textbf{Example:} \textcolor{black}{Considering the case that GPT has applied UC1 to identify the missed essential actor `Gold Customer'. Accordingly, GPT deducted points for missing the links between `Gold Customer' and its use cases, by following UC4 to evaluate the same use case diagram. However, these missing links were not identified by human experts.}
\end{mdframed} 
    
\end{itemize}

\textbf{Findings:} \textcolor{black}{In assessing UML use case diagrams, the best performance of GPT is to grade the relationship between actors. \textit{`Overstrictness'} and \textit{`Wrong Identification'} are two main factors that undermine the grading consistency of GPT with human experts. The former mainly occurred in identifying actors and use cases with UC1 and UC2 respectively, and the latter was the most reported in evaluating the relationships between identified actors and use cases with UC4. }

\subsubsection{Competency of GPT in assessing class diagram}
\label{competency4C}
Table~\ref{tab:discc} summarizes the occurrences of different evaluation discrepancies when using each CC to grade \textcolor{black}{40} UML class diagrams. \textcolor{black}{More specifically, when GPT evaluates UML classes with CC1, CC2, and CC3, \textit{`Misunderstanding'} is the most frequent evaluation discrepancy that occurs in 38, 24, and 25 of 40 students' class diagrams, respectively. However, GPT often deducts points due to the wrong identification of relationships between classes with CC4, which is reported in 19 of the 40 class diagrams. Note that the subtotals of the occurrence of evaluation discrepancies using CC1 and CC4 are 49 and 47, respectively. This indicates that more than one type of evaluation discrepancies are more likely to occur when GPT applies CC1 and CC4 to evaluate class diagrams created by the 40 students. }

\begin{table*}[h]
\scriptsize
    \caption{\textcolor{black}{Occurrence of different types of discrepancies in the assessment of class diagrams}}
    \label{tab:discc}
    \centering   
    \begin{tabular}{l c c c c }
    \hline
    \textbf{Type of Evaluation Discrepancies (Occur.)} & \textbf{CC1} & \textbf{CC2} & \textbf{CC3} & \textbf{CC4} \\ \hline
    Perfect Match (33) & 2 & 13 & 11 & 7 \\ \hline
    Wrong Identification (23) & 0 & 2 & 2 & 19 \\ \hline
    Overstrictness (20) & 9 & 0 & 0 & 11 \\ \hline
    Misunderstanding (91) & 38 & 24 & 25 & 4 \\ \hline
    Reasonable Identification (10) & 0 & 2 & 2 & 6 \\ \hline
    \textit{Total:} & 49 & 41 & 40 & 47 \\ \hline
    \end{tabular}
\end{table*}

\begin{figure}[h]
\scriptsize
\centering
\includegraphics[width=1.0\textwidth]{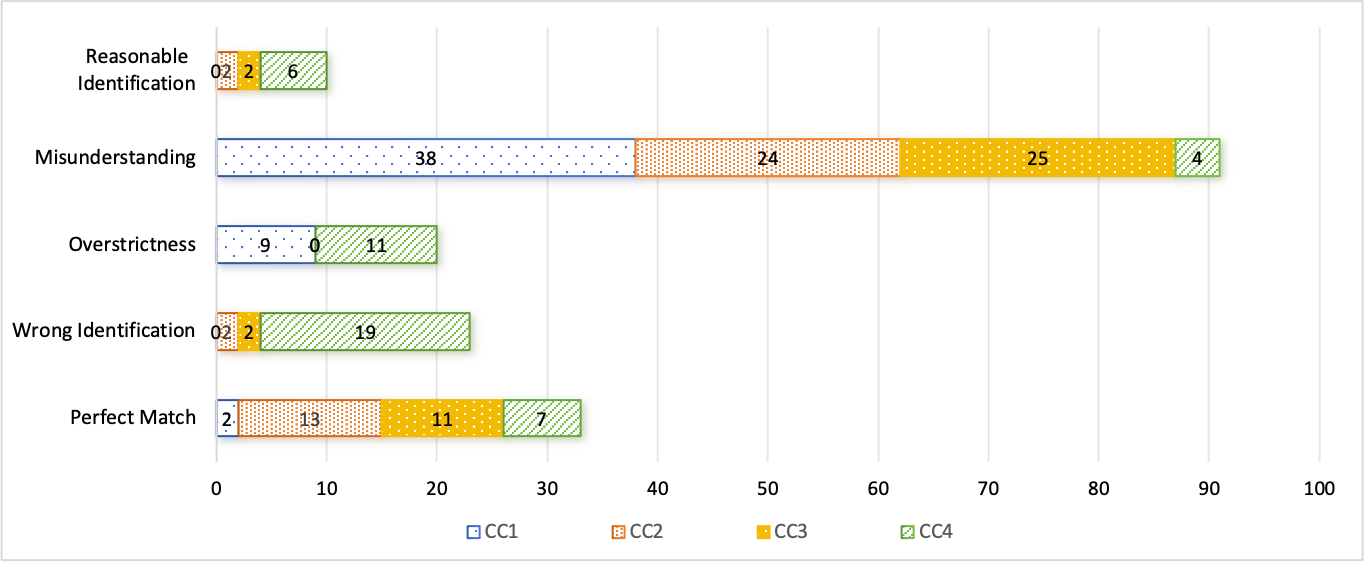}
\caption{\textcolor{black}{Occurrence distribution of the five discrepancy types in grading 40 UML class diagrams with four CCs}}
\label{fig:ccsta}
\end{figure}

\textcolor{black}{Figure~\ref{fig:ccsta} zooms in on the occurrence distribution of the five types of discrepancies when human experts and GPT independently use four CCs to grade 40 UML class diagrams.} The observations and the corresponding examples are listed as follows.

\begin{itemize}
 \item \textbf{Perfect Match} 
   \textcolor{black}{occurs 33 in times evaluating UML class diagrams, as shown in Table~\ref{tab:discc}. Especially, the last bar of Figure~\ref{fig:ccsta} presents that 39.4\% of \textit{`Perfect Match'} (13 of 33 occurrences) is reported in the use of CC2 and 33.3\% is in the use of CC3. This indicates that relatively, GPT performs well in recognizing attributes and operations of the identified classes in our case study. However, GPT is not good at understanding and using CC1 to identify classes.}

\begin{mdframed}
    \textbf{Example:} \textcolor{black}{During the evaluation based on CC2 and CC3, both human scorers and GPT found a certain amount of reasonable attributes and operations of the identified classes.}
\end{mdframed}

\item \textbf{Wrong Identification}   
    \textcolor{black}{is evaluation discrepancy with lower occurrence in the GPT-based assessment of UML class diagrams, with 23 occurrences in Table~\ref{tab:discc}. Specifically, the fourth bar in Figure~\ref{fig:ccsta} shows 82.6\% of \textit{`Wrong Identification'} occurred in the evaluation based on CC4 (19 of 23 occurrences). This type of discrepancy reflects the limitations of GPT in correctly identifying the relationships between classes. Another observation is that there are no \textit{`Wrong Identification'} when GPT employs CC1 for class identification.}

 \begin{mdframed}
     \textbf{Example:} \textcolor{black}{In the evaluation based on CC4, the generalization relationship between `Golden Customer' and `Customer' did exist in the UML class diagram and presented with the correct UML notation, but GPT failed to identify it and then deducted points.}
 \end{mdframed}

\item \textbf{Overstrictness} \textcolor{black}{is also the evaluation discrepancy with lower occurrence in the GPT-based assessment of UML class diagrams, with 20 occurrences in Table~\ref{tab:discc}. As the third bar of Figure~\ref{fig:ccsta} shows, all the \textit{`Overstrictness'} are reported in the evaluation based on CC1 and CC4, accounting for 45\% and 55\% in identifying the essential classes and the relationships between the classes in the UML class diagram, respectively.} 
    
\begin{mdframed}
   \textbf{Example:} \textcolor{black}{Considering the evaluation based on CC1, the human experts regarded the class `Notification' as being partially equivalent to the class `BackOrderNotifier' in the case study. However, the evaluation conducted by GPT was rigorous, resulting in points deduction. Another example is the composition/association relationship between `Order' and `Product'. Human experts treated this relationship as the identified one with an inappropriate type, and as a result, they deducted 0.1 points. However, GPT deducted 0.2 points because it detected the omission of this aggregation relationship.}
\end{mdframed}

\item \textbf{Misunderstanding} \textcolor{black}{is the most prominent type of discrepancy reported in the GPT-based assessment of UML class diagrams, with 91 occurrences in Table~\ref{tab:discc}. As the second bar of Figure~\ref{fig:ccsta} shows, GPT misunderstood CC1 in evaluating 95\% of the 40 class diagrams. This reflects the main limitation GPT in correctly identifying a certain number of essential classes. Meanwhile, GPT misunderstood CC2 and CC3 in grading a certain number of reasonable attributes and operations of the identified classes in 24 and 25 UML class diagrams, respectively.}

 \begin{mdframed}
    \textbf{Example:} \textcolor{black}{In the case of a UML class diagram containing five classes, four of them are included in the ten suggested classes. When grading this class diagram with CC1, human experts deducted 0.8 points for missing four suggested classes. However, GPT checked the ten suggested classes and found that six of them were not included in this class diagram. As a result, GPT gave 0 points to this class diagram after evaluating based on CC1.}
\end{mdframed}

    \item\textbf{\textcolor{black}{Reasonable Identification}}
    \textcolor{black}{is the discrepancy type with the lowest occurrence reported in the GPT-based assessment of UML class diagrams, with 10 occurrences in Table~\ref{tab:discc}. The first bar of Figure~\ref{fig:ccsta} represents that \textit{`Reasonable Identification'} effectively pinpointed some attributes and operations of classes as well as the relationships between classes that were overlooked by human experts but essential for inclusion in the class diagrams. Specifically, GPT's understanding and use of CC2, CC3, and CC4 provided supplementary insights to the evaluation of two, two, and six class diagrams, respectively.}
    
\begin{mdframed}
    \textbf{Example:} \textcolor{black}{In the case that GPT applied CC4 to identify the missed aggregation relationship between `Order' and `Invoice'. GPT deducted 0.2 points for the omission of this aggregation relationship, but human experts did not deduct points for this omission.}
\end{mdframed}

\end{itemize}

\textbf{Findings:} In the assessment of UML class diagrams, \textcolor{black}{\textit{`Misunderstanding'} is the main factor that causes the grading difference between GPT and human experts. Moreover, it occurs frequently when applying CC1, CC2, and CC3 to identify essential classes as well as their attributes and operations.}

\subsubsection{Competency of GPT in assessing sequence diagram}
\label{competency4S}
Table~\ref{tab:dissc} summarizes the occurrences of different types of evaluation discrepancies when GPT used each of the three SCs to evaluate \textcolor{black}{40} UML sequence diagrams. \textcolor{black}{More specifically, when GPT evaluates UML sequence diagrams with SC1, \textit{`Misunderstanding'} and \textit{`Overstrictness'} are the two most frequent evaluation discrepancies, occurring in 36 and 18 of 40 students' sequence diagrams, respectively. 
Regarding GPT's evaluation with SC2, \textit{`Misunderstanding'} is also the most frequent evaluation discrepancy with 19 occurrences. The second and third common assessment discrepancies reported in using SC2 are \textit{`Wrong Identification'} and \textit{`Overstrictness'}.
GPT often deducts points due to wrong identifications of the order of messages, which occurs in 21 out of the 40 sequence diagrams.
Note that the subtotals of the occurrence of evaluation discrepancies using SC1 and SC2 are 64 and 56, respectively. This indicates that more than one type of evaluation discrepancies are more likely to occur when GPT applies SC1 and SC2 to evaluate sequence diagrams created by the 40 students. }

\begin{table*}[h]
\scriptsize
    \caption{\textcolor{black}{Occurrence of different types of discrepancies in the assessment of sequence diagrams}}
    \label{tab:dissc}
    \centering   
    \begin{tabular}{l c c c }
    \hline
    \textbf{Type of Evaluation Discrepancies (Occur.)} & \textbf{SC1} & \textbf{SC2} & \textbf{SC3}  \\ \hline
    Perfect Match (9) & 0 & 0 & 9  \\ \hline
    Wrong Identification (39) & 2 & 16 & 21  \\ \hline
    Overstrictness (35) & 18 & 14 & 3 \\ \hline
    Misunderstanding (56) & 36 & 19 & 1\\ \hline
    Reasonable Identification (21) & 8 & 7  & 6 \\ \hline
    \textit{Total:} & 64 & 56 & 40 \\ \hline
    \end{tabular}
\end{table*}

\begin{figure}[h]
\scriptsize
\centering
\includegraphics[width=0.9\textwidth, height=0.28\textheight]{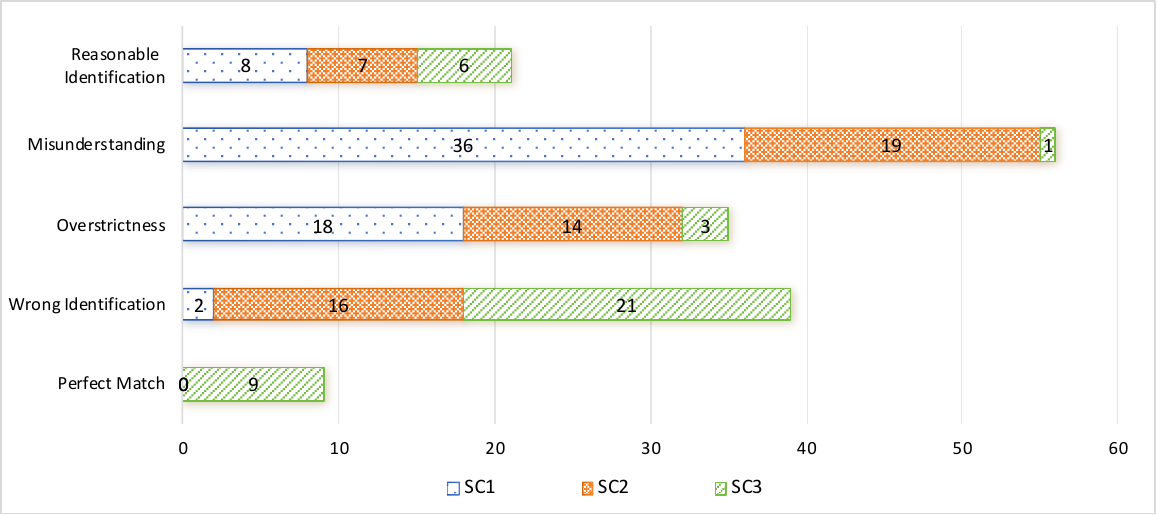}
\caption{\textcolor{black}{Occurrence distribution of the five discrepancy types in grading 40 UML sequence diagrams with three SCs}}
\label{fig:sqsta}
\end{figure}

\textcolor{black}{Figure~\ref{fig:sqsta} zooms in on the occurrence distribution of the five types of discrepancies when human experts and GPT independently use three SCs to grade 40 UML sequence diagrams.} The observations and the corresponding examples are listed as follows.

\begin{itemize}
 \item \textbf{Perfect Match} 
   \textcolor{black}{occurs 9 times in evaluating 40 UML sequence diagrams, as shown in Table~\ref{tab:dissc}. Especially, the last bar of Figure~\ref{fig:sqsta} presents that all the \textit{`Perfect Match'} is reported in the use of SC3. This indicates that relatively, GPT performs well in recognizing and evaluating orders of messages in the sequence diagrams.}

\begin{mdframed}
    \textbf{Example:} \textcolor{black}{Both human scorers and GPT found some unreasonable orders of messages in the UML sequence diagrams to be evaluated, according to the description of the case study.}
\end{mdframed}

\item \textbf{Wrong Identification}   
    \textcolor{black}{is evaluation discrepancy with the second highest occurrence in GPT-based assessment of UML sequence diagrams, with 39 occurrences in Table~\ref{tab:dissc}. 
    Specifically, the fourth bar in Figure~\ref{fig:sqsta} shows that 53.8\% and 41\% of \textit{`Wrong identification'} occurred in the evaluation based on SC3 (21 of 39 occurrences) and SC2 (16 of 39 occurrences), respectively. This type of discrepancy reflects the limitations of GPT in correctly identifying and grading messages and their orders in the UML sequence diagram.}

 \begin{mdframed}
     \textbf{Example:} \textcolor{black}{In the evaluation based on SC2, GPT failed to deduct points for the messages without required parameters. In the evaluation based on SC3, GPT deducted points because `the sequence of confirming shipping/billing info and submitting the order is not clearly separated'. However, these messages were presented separately and organized in a reasonable order.}
 \end{mdframed}

\item \textbf{Overstrictness} \textcolor{black}{is the evaluation discrepancy with the third highest frequency in the GPT-based assessment of UML sequence diagrams, with 35 occurrences in Table~\ref{tab:dissc}. As the third bar of Figure~\ref{fig:sqsta} shows, \textit{`Overstrictness'} are mainly reported in the evaluation based on SC1 and SC2, accounting for 51.4\% and 40\% in identifying objects and the messages between them in the UML sequence diagram, respectively.} 
    
\begin{mdframed}
   \textbf{Example:} \textcolor{black}{Considering the evaluation based on SC1, the human experts considered the identified object `Order Processing System' to be equivalent to  the suggested answer `OrderProcessingManager'. However, GPT deducted points for the reason `Missing essential object: OrderProcessingManager (or equivalent) is not present; `Order Processing System' is used but does not clearly represent the manager/controller object.'}
\end{mdframed}

\item \textbf{Misunderstanding} \textcolor{black}{is the most prominent type of discrepancy reported in the GPT-based assessment of UML sequence diagrams, with 56 occurrences in Table~\ref{tab:dissc}. As the second bar of Figure~\ref{fig:sqsta} shows, GPT misunderstood SC1 and SC2 in evaluating 70.6\% and 37.3\% of the 40 sequence diagrams, respectively. This reflects the main limitation GPT in correctly identifying a certain number of essential objects and reasonable messages between objects.}

 \begin{mdframed}
    \textbf{Example:} \textcolor{black}{In the case of a UML sequence diagram containing five objects, four of them are included in the five suggested objects. When grading this sequence diagram with SC1, human experts gave a full score. However, GPT checked the five suggested classes and found that one of them were not included in this sequence diagram. As a result, GPT deducted 0.2 points after evaluating this sequence diagram based on SC1.}
\end{mdframed}

    \item\textbf{\textcolor{black}{Reasonable Identification}}
    \textcolor{black}{is the discrepancy type with the lowest occurrence reported in the GPT-based assessment of UML sequence diagrams, with 9 occurrences in Table~\ref{tab:dissc}. The first bar of Figure~\ref{fig:sqsta} represents that all the \textit{`Reasonable Identification'}-typed discrepancy are reported when GPT employed SC3 to evaluate the order of identified messages.}
    
\begin{mdframed}
    \textbf{Example:} \textcolor{black}{In the case that GPT applied SC3 to identify and grade the order of messages between the identified objects, GPT deducted points for some unreasonable orders, but human experts failed to identify these irrationalities in the orders.}
\end{mdframed}

\end{itemize}

\textbf{Findings:} \textcolor{black}{In the assessment of UML sequence diagrams, \textit{`Misunderstanding'}, \textit{`Wrong Identification'}, and \textit{`Overstrictness'} are three main factors that cause the grading difference between GPT and human experts. Specifically, \textit{`Misunderstanding'} and \textit{`Overstrictness'} occur frequently when GPT employs SC1 and SC2 to identify objects and messages between objects in sue, and \textit{`Wrong Identification'} is usually reported in the evaluation of messages and their orders with SC2 and SC3.}

\begin{figure}[h]
\scriptsize
\centering
\includegraphics[width=1.0\textwidth]{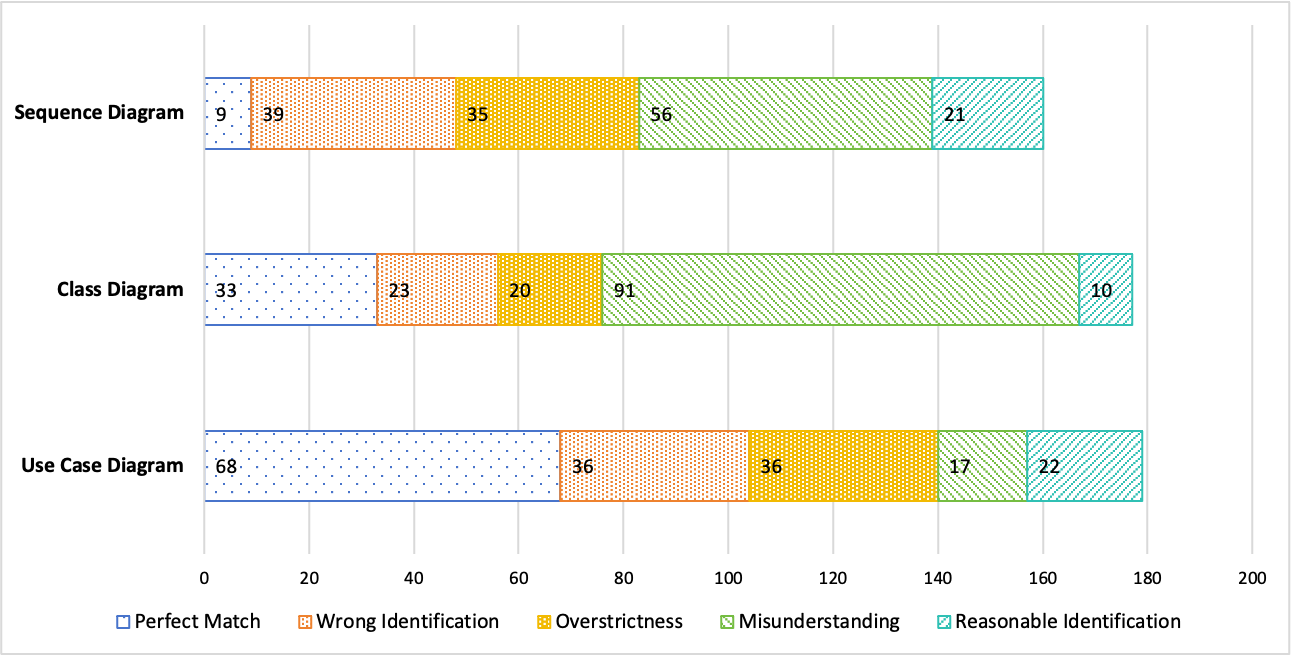}
%\caption{\textcolor{black}{Evaluation discrepancy of GPT in assessing three types of UML diagrams}}
\caption{\textcolor{black}{Occurrence distribution of evaluation discrepancy over three types of UML diagrams}}
\label{fig:rq2}
\end{figure}

\textcolor{black}{Figure~\ref{fig:rq2} compares the percentages of five evaluation discrepancies when GPT is used to evaluate three types of UML diagrams. We found that the percentage of \textit{`Perfect Match'}, \textit{`Overstrictness'}, and \textit{`Wrong Identification'} in evaluating UML use case diagrams is much higher than their percentages in evaluating UML class and sequence diagrams. \textit{`Misunderstanding'} is the primary evaluation discrepancy reported in the GPT-based evaluation of UML class and sequence diagrams. \textit{`Reasonable Identification'} is reported when GPT grades all three types of UML diagrams.}

\begin{center}
\fcolorbox{black}{gray!10}{\parbox{0.97\linewidth}{\textbf{Key Findings of RQ2:} \textcolor{black}{First, GPT performed better in grading UML use case diagrams than UML class and sequence diagrams. Second, when evaluating UML class and sequence diagrams, GPT usually misunderstood the evaluation criteria. Whereas, it seems to be more difficult for GPT to correctly identify the specified modeling elements in the UML use case and sequence diagrams. In addition, in the evaluation of three types of UML diagrams, GPT can effectively identify some reasonable modeling elements that are overlooked by human experts.} }}
\end{center}

\section{Discussion}
In this section, we discuss the experimental results presented in Section~\ref{sec:results} and analyze the potential impacts and significance that GPT may bring to the education of SE in universities. 

\subsection{Interpretation of RQ1 Results}
The answers to RQ1 reveal that by feeding GPT the role (e.g., a computer science lecturer in this paper) it will play in the specified task and detailed evaluation criteria, the scores and the corresponding reasons for deducting points can be automatically given to the specified UML diagrams. Regarding 40 student reports, the results in Figure~\ref{fig:scorediff1} indicate that the score difference between human experts and GPT span a range of -0.25 to +3.75. Meanwhile, as observed in Table~\ref{tab:diffscore}, although the scores provided by GPT were lower than those provided by human experts when evaluating more than 80\% of the UML diagrams created by 40 students, each type of UML diagrams can get the same scores from GPT and human scorers. This indicates that GPT can work as an AI scorer to grade course assignments in the UML modeling course with well-designed and structured prompts, but it cannot fully substitute for human teachers or teaching assistants yet.

In addition, as shown in Table~\ref{tab:diffscore} and Figure~\ref{fig:scorediff2}, the score differences induced by GPT exhibit distinct fluctuations when evaluating different types of UML diagrams. This reveals that the performance of GPT is different in evaluating different types of UML diagram. One possible reason is that the characteristics of the modeling elements and the complexity of the model structure differ from one type of UML diagram to another. This also motivates a deeper analysis of how GPT handles different modeling constructs in various types of UML diagram, which is what we intend to explore through RQ2.

\subsection{Interpretation of RQ2 Results}
The findings of RQ2 further elaborate on the differences in both the subscores and the total scores of each UML diagram created by the students reported in the answer to RQ1. Referring to the scores provided by human experts, these differences reveal GPT's competency in using specified evaluation criteria to assess various types of UML diagrams. 

Based on the scores and the reasons for deducting points, we first identified and summarized five types of evaluation discrepancies, i.e., \textit{Overstrictness}, \textit{Wrong Identification}, \textit{Reasonable Identification}, and \textit{Perfect Match}, to assess GPT's performance in understanding and applying the proposed evaluation criteria to grade UML use case, class, and sequence diagrams. More specifically, \textit{Misunderstanding}, \textit{Overstrictness}, and \textit{Wrong Identification} characterize the weakness of GPT in the task of evaluating UML diagrams, and \textit{Reasonable Identification} and \textit{Perfect Match} represent the strength of GPT in this task. These five discrepancy types pave the way for systematically analyzing both the limitations and competencies of GPT in automated UML diagram evaluation, offering valuable insights for future improvements in AI-assisted software modeling assessments.

Section~\ref{competency4UC} reports GPT's performance in understanding and using four UCs in Table~\ref{tab:uccriteria} to grade 40 UML use case diagrams. The findings in Table~\ref{tab:disuc} indicate that \textit{`Perfect Match'} is the most frequent type reported in the GPT-based evaluation of use case diagrams, accounting for 68 occurrences. This implies that GPT demonstrates particular proficiency in the automated evaluation of UML use case diagrams. Meanwhile, either \textit{`Wrong Identification'} or \textit{`Overstrictness'} occurred 36 times when GPT evaluated 40 use case diagrams. This implies that when evaluating use case diagrams, the main limitations of GPT are the wrong identification of specified modeling elements and excessively rigorous grading. Considering the modeling constructs use case diagrams, GPT performed too rigorously in identifying actors and use cases with UC1 and UC2, accounting for 91.6\% of reported \textit{`Overstrictness'}, as listed in Table~\ref{tab:disuc}. The main reason is that actors and use cases in these use case diagrams are usually named in Chinese or with English phrases that are semantically similar to the suggested answers. This indicates a limitation of GPT in understanding and processing Chinese and synonymous terms or equivalent phrases in English. Regarding the structure of use case diagram, GPT employs UC3 to evaluate the relationship between actors and reported 37 occurrences of \textit{`Perfect Match'}. This revealed the strength of GPT in identifying the relationship between two actors in our case study. The main reason is that there is only one suggested relationship between two actors in our suggested answer, i.e., the generalization between `Customer' and `Gold Customer'. The suggested answer is relatively simple, so it is easy for GPT to understand and evaluate correctly, as human experts did. As for the relationships between use cases, GPT employed UC4 and triggered 24 occurrences of \textit{`Wrong Identification'}, which may be due to two reasons. One possible reason is the influence of wrongly identified use cases. The other reason is GPT's weak capability in parsing and identifying solid lines between actors and use cases in \texttt{.png} files. 

Section~\ref{competency4C} reported GPT's performance in understanding and using four CCs in Table~\ref{tab:classcriteria} to grade 40 UML class diagrams. The findings in Table~\ref{tab:discc} indicate that \textit{`Misunderstanding'} is the most frequent type reported in the GPT-based evaluation of UML class diagrams, accounting for 91 occurrences. This reveals the primary limitation of GPT in grading UML class diagrams. Generally, the class with appropriate attributes and operations is treated as the essential modeling construct in UML class diagram. However, 95.6\% of the occurrences of \textit{`Misunderstanding'} were reported in the evaluation of class diagrams with CC1, CC2, and CC3. 
The main reason could be that the scoring criteria listed in the last column of Table~\ref{tab:classcriteria} appear to be ambiguous and lack sufficient clarity for GPT, making it difficult for GPT to understand and use them correctly. These not only expose GPT's weakness in evaluating UML class diagrams but also suggest the strategies to improve GPT's capability in this task. 
Another possible reason is that the grading of attributes and operations is conducted after the corresponding class is confirmed, which follows the regular steps conducted by human experts, i.e., to identify the class before its attributes and operations. If GPT misunderstands CC1 and cannot identify required number of essential classes, it is likely to propagate errors throughout the evaluation process of attribute and operation based on CC2 and CC3. 
The complex structure of a UML class diagram comes mainly from the number of classes and different kinds of relationships between classes. As shown in Figure~\ref{fig:ccsta}, 82.6\% of 23 occurrences of \textit{`Wrong Identification'} were reported when GPT used CC4 to evaluate the relationships between classes. This is probably because of GPT's weak capability in parsing and identifying different types of line between classes in \texttt{.png} files. Another possible reason is the influence of classes that cannot be identified appropriately and correctly before GPT uses CC4. Similarly, these limitations have been evident in GPT's performance in evaluating UML use case diagrams.
%Another possible reason could be that as the first and foremost element in class diagrams, classes represent an abstraction formed by comprehending and distilling key information in relevant text, rather than simply reusing the nouns explicitly listed in the sentences. 
%\textcolor{black}{Another potential reason is that the definition of a class, along with its attributes and operations, is inherently subjective. This subjectivity gives rise to a wide variety of naming conventions for classes that represent the same concept and operations that fulfill identical responsibilities. However, GPT may struggle with processing synonymous or similar terms, and this limitation is also evident in its performance in evaluating UML use case diagrams. 

\textcolor{black}{Section~\ref{competency4S} reported GPT's performance in understanding and using three SCs in Table~\ref{tab:seqcriteria} to grade 40 UML sequence diagrams. The findings in Table~\ref{tab:dissc} indicate that similar to the UML class diagram, \textit{`Misunderstanding'} is also the most frequent type reported in the GPT-based evaluation of UML sequence diagrams, accounting for 56 occurrences. This reveals the primary limitation of GPT in grading UML sequence diagrams. Generally speaking, objects are the essential modeling construct in the UML sequence diagram, and the structure of a sequence diagram depends on the number of messages from one object to another and the order of these messages. However, the results in Figure~\ref{fig:sqsta} show that 98.2\% of 56 occurrences of \textit{`Misunderstanding'} reported when GPT used SC1 and SC2 to identify objects and messages. Similarly to UML class diagrams, the main reason could be that the scoring criteria listed in the last column of Table~\ref{tab:seqcriteria} seem to lack sufficient clarity for GPT, making it difficult for GPT to understand and use them correctly. These not only expose GPT's weakness in evaluating UML sequence diagrams but also suggest strategies to improve GPT's capability in this task. 
In GPT-based assessment with SC1 and SC2, another commonly observed type of evaluation discrepancy is \textit{`Overstrictness'}, as Figure~\ref{fig:sqsta} shows. One possible reason is that objects and messages in these sequence diagrams are usually named in Chinese or with English phrases that are semantically similar to the suggested answers. Similarly to GPT-based evaluation of UML use diagrams, this indicates a limitation of GPT in understanding and processing Chinese and synonymous terms or equivalent phrases in English. Another reason is the complexity and dynamics of sequence diagrams, which brings more flexibility to the evaluation process of sequence diagrams. Compared to human experts, however, GPT often struggles to tolerate alternative reasonable answers.
In addition, \textit{`Wrong Identification'} is the second most commonly observed type in the evaluation of sequence diagrams performed by GPT, especially in the evaluation based on SC2 and SC3. This is partially because of GPT's weak capability in parsing and identifying the messages written in Chinese and the UML notation for messages in \texttt{.png} files.}

\textcolor{black}{Finally, it is interesting to observe that during the evaluation process carried out by GPT, \textit{`Reasonable Identification'} occurs in grading UML use case diagrams with three UCs (i.e., UC1, UC2, and UC4), class diagrams with three CCs (i.e., CC2, CC3, and CC4), and sequence diagrams with all three SCs. This not only reveals the strength of GPT in evaluating UML diagrams but also highlights the valuable supplementary insights GPT provides in SE education.}

\subsection{Implications}
The answers to RQ1 and RQ2 reveal the feasibility and performance of GPT in evaluating and giving scores to UML use case diagrams, class diagrams, and sequence diagrams. Although the results expose some limitations of GPT in assessing UML diagrams, it is promising to employ GPT as AI educators in the field of SE education. 

\textbf{For teachers:} The research findings inspire new directions for SE education, especially teaching activities in the era of AI. First, the employment of GPT makes it possible to review and grade students' course projects or assignments automatically and efficiently. This allows teachers to devote more time and energy to teaching courses. Second, GPT can mitigate biases that may arise from human experts who are fatigued after reviewing a large number of UML diagrams. This ensures the objectivity of the grading process and the fairness of scores to some degree. Third, \textcolor{black}{since the UML diagrams to be evaluated are different, the reasons for the point deduction provided by GPT are personalized feedback to the specified UML diagrams.} Therefore, these reasons can serve as valuable information to help teachers understand students' learning situations and identify their weaknesses. Based on such information, teachers can tailor their teaching plans to better meet the needs of students. 

\textbf{For students:} Instead of manually reviewing their UML diagrams, GPT enables students to quickly receive scores and personalized feedback. This enables students to efficiently and effectively understand their learning status for improvement. However, due to the weaknesses and limitations of GPT reported in Section~\ref{RQ2}, these students need to strengthen their ability to identify and correct the errors made by GPT. Moreover, this work opens the door for students to improve and polish their UML diagrams before submitting. In this situation, GPT can help them create UML diagrams with higher quality and then get higher final scores for relevant courses. 

\textbf{For researchers:} This study paves the way to apply other generative AI techniques to aid in evaluating the three types of UML diagrams created by the students. Therefore, researchers are expected to explore the performance of diverse generative AI techniques in this task or other similar tasks in SE education. In addition, more research is necessary to improve GPT's proficiency in assessing UML use case diagrams, class diagrams, and sequence diagrams. For example, prompt engineering can be utilized to refine and polish the prompts to evaluate these UML diagrams. \textcolor{black}{Further improvements on the grading specifications of the proposed evaluation criteria are also expected to mitigate the \textit{`Misunderstanding} and \textit{`Overstrictness'} issues that GPT brings to the automatic evaluation of UML diagrams.}

\section{Threats to Validity}
\label{limitations}
This section follows the guidelines in ~\citep{shull2008guide} and ~\citep{wohlin2012experimentation} to evaluate the potential threats to the validity of our research results.

\textbf{Construct validity:} Due to the inherent randomness exhibited by current generative AI techniques, including GPT used in this study, identical inputs (e.g., the picture of UML diagrams) may lead to different outputs (e.g., scores of the input UML diagrams). Therefore, the performance of GPT on the evaluation of UML diagrams cannot be fully controlled with precision by formatting the prompts. Meanwhile, specific training of GPT seems to help improve the performance  of GPT in this evaluation task. In this paper, however, there is no information on the knowledge base required for proper training of GPT before it is used to evaluate UML diagrams. In the future, modeling experience of human experts and UML specifications could be imported as a knowledge base for training GPT to improve its performance in evaluating UML diagrams. \textcolor{black}{In addition, the evaluation criteria proposed for the three types of UML diagrams are based on the answers suggested in the supplementary teaching material of our course. On one hand, since complex modeling constructs, such as relationships between use cases, multiplicities adhering to the relationships between classes, control flow and `ref' frame in sequence diagrams, are not represented in the suggested UML diagrams, the grading criteria of these modeling constructs are not taken into account or are explicitly simplified as `No points are deducted'. On the other hand, considering the complexity and subjectiveness in the process of evaluating some modeling elements (e.g., class attributes and operations), some evaluation criteria (e.g., CC2 and CC3) are proposed to enable both human experts and GPT to accommodate diverse design choices while maintaining a shared standard of adequacy. These may impact the analysis and assessment of GPT's capabilities. In the next step, to improve the evaluation criteria (e.g., by removing ambiguous expressions) and refine the prompt (e.g., by introducing Chain of Thought) for GPT may partially mitigate this threat.}
\textcolor{black}{Another threat arises from the fact that the 40 undergraduate students were permitted to use LLMs (e.g., ChatGPT) when creating their UML diagrams, as mentioned in Section~\ref{sec:data}. Since this work evaluates the performance of GPT in scoring these diagrams, there is a risk of assessment bias that GPT may be more inclined to better understand diagrams generated by or with the help of similar and/or the same version of GPT. This may limit the generalizability of the findings to scenarios where UML diagrams are created by humans independently or (partially) by other LLMs.}

\textbf{Internal validity:} Since the analysis and evaluation of the UML diagrams are based on the evaluation criteria proposed for the three types of UML diagrams and the suggested answers, the threats to internal validity introduced by the subjectivity and bias of the scorers are minimized. In addition, the scoring method adopted in this study may not fully capture the accuracy and completeness of UML use case diagrams, class diagrams, and sequence diagrams. In this paper, the scoring details of each evaluation criterion were first defined based on the teaching and grading experience of the authors and then polished by pilot evaluation and discussion between the authors. This measure can partially reduce this threat. 

\textbf{External validity:} There is an external threat to validity with respect to the limited generalizability of the research findings because GPT-4o is the only LLM used in this study. The results may vary if \textcolor{black}{other versions of GPT or even other LLMs} are employed to grade the three types of UML models. Thus, the performance of more generative AI models in evaluating UML diagrams remains to be studied in our future work. In addition, 40 project reports used in this paper come from two student groups of the same university. This partially mitigates the threat caused by the sample size. However, the UML diagrams created by students from other universities or practitioners are expected to help improve the external validity of the research results. 

\textbf{Conclusion validity:} Due to time and resource limitations, the research data of this work, i.e., the UML diagrams to be graded, were created by 40 undergraduate students at our university and collected from their course project reports. The research findings may be affected by the import of the UML diagrams created by the practical modelers, varying in educational background and modeling experience, for the same case study used in this study. Another threat to the conclusion validity could be caused by the UML diagrams created for different application domains or modeled with different levels of complexity and size, which will be further investigated in our future work.

\section{Conclusions and Future Work}
\label{conclusions}
In this exploratory study, we quantitatively explored the capability of GPT in evaluating and grading three types of UML diagrams, i.e., use case diagrams, class diagrams, and sequence diagrams. More specifically, we first collected UML use case diagrams, class diagrams, and sequence diagrams provided in the course project reports of 40 undergraduate students. Then, we defined the evaluation criteria and the grading details for these three types of UML diagrams, respectively. These evaluation criteria not only guided the grading process and results conducted by human experts but also were integrated into the prompt feeding to GPT to get grading scores of input UML diagrams. By comparing and analyzing the scores provided by human experts and GPT, \textcolor{black}{it is concluded that GPT can be used in evaluating specified types of UML diagrams (e.g., use case diagrams, class diagrams, and sequence diagrams), but it cannot substitute for human experts yet.} These score differences between GPT and human experts were further studied by analyzing the evaluation discrepancies that GPT brought by using each evaluation criterion. We found that GPT often evaluates the UML diagrams too rigidly and its performance notably differs in the assessment of three types of UML diagrams. 
 
The next steps of our work are the following: \textcolor{black}{(1) to mitigate the \textit{`Misunderstanding'} and \textit{`Overstrictness'} issues reported in GPT-based evaluation by refining the proposed evaluation criteria or to give proper training for GPT with required professional knowledge base,} (2) to expand the sample size of the study by collecting UML diagrams created by students from other universities or practical modelers, (3) to extend the experiments to evaluate more types of UML diagrams, such as UML state diagram and activity diagram, \textcolor{black}{and to verify the traceability between specified UML diagrams,} and (4) to evaluate the capabilities of other LLMs in grading UML diagrams, as well as in requirements analysis and software modeling tasks in SE education.

\section*{Data availability}
The replication package for this work has been made available at \url{https://github.com/RorschachXR/GPT4UMLEval}.

%\appendix
%\section*{Appendix}
%A brief description for each selected MPL OSS Apache project is shown in Table \ref{table-project-description}, 

\section*{Acknowledgements}
This work has been partially supported by the National Natural Science Foundation of China (NSFC) with Grant No. 62172311.

%\bibliographystyle{cas-model2-names}

\bibliography{references}

%\end{frontmatter}
%\end{sloppypar}
\end{document}